\documentclass[useAMS,usenatbib]{mn2e}

\usepackage{txfonts}
\usepackage{graphicx}

\title[The Outer Halo of the Nearest Giant Elliptical]{The Outer Halo
  of the Nearest Giant Elliptical: A VLT/VIMOS Survey of the Resolved
  Stellar Populations in Centaurus~A to 85~kpc\thanks{Based on data
    collected at the European Southern Observatory, Paranal, Chile,
    within the observing Programme 074.B-0741.}}
\author[D. Crnojevi\'c et al.]{D. Crnojevi\'c$^{1}$\thanks{Email: dc@roe.ac.uk}, A. M. N. Ferguson$^{1}$, M. J. Irwin$^{2}$, E. J. Bernard$^{1}$, N. Arimoto$^{3,4}$, \newauthor P. Jablonka$^{5,6}$, C. Kobayashi$^{7,8}$ \\
  $^{1}$Institute for Astronomy, University of Edinburgh, Royal Observatory, Blackford Hill, EH9 3HJ Edinburgh, UK\\
  $^{2}$Institute of Astronomy, Madingley Road, CB3 0HA Cambridge, UK\\
  $^{3}$Subaru Telescope, National Astronomical Observatory of Japan, 650 North A'ohoku Place, Hilo, Hawaii 96720, USA\\
  $^{4}$Department of Astronomical Science, Graduate University for Advanced Studies, Mitaka, Tokyo 181-8588, Japan\\
  $^{5}$Laboratoire d’Astrophysique, Ecole Polytechnique F\'ed\'erale de Lausanne, Observatoire, CH-1290 Sauverny, Switzerland\\
  $^{6}$GEPI, Observatoire de Paris, CNRS UMR 8111, Universit\'e Paris Diderot, F-92125 Meudon Cedex, France\\
  $^{7}$School of Physics, Astronomy and Mathematics, University of Hertfordshire, AL10 9AB Hatfield, UK\\
  $^{8}$Research School of Astronomy and Astrophysics, The Australian
  National University, Cotter Road, Weston, ACT 2611, Australia}

\begin{document}

\date{Accepted 2013 March 18. Received 2013 March 6; in original form 2013 January 16}

\pagerange{\pageref{firstpage}--\pageref{lastpage}} \pubyear{}

\maketitle

\label{firstpage}

\begin{abstract}

  We present the first survey of resolved stellar populations in
  the remote outer halo of our nearest giant elliptical (gE),
  Centaurus~A (D$=3.8$~Mpc). Using the VIMOS/VLT optical camera, we
  obtained deep photometry for four fields along the major and minor
  axes at projected elliptical radii of $\sim30-85$~kpc (corresponding
  to $\sim5-14 R_{\rm eff}$).  We use resolved star counts to map the
  spatial and colour distribution of red giant branch (RGB) stars down
  to $\sim2$ magnitudes below the RGB tip.  We detect an extended halo
  out to the furthermost elliptical radius probed ($\sim85$~kpc or
  $\sim14 R_{\rm eff}$), demonstrating the vast extent of this system.
  We detect localised substructure in these parts, visible in both
  (old) RGB and (intermediate-age) luminous asymptotic giant branch
  stars, and there is some evidence that the outer halo becomes more
  elliptical and has a shallower surface brightness profile.  We
  derive photometric metallicity distribution functions for halo RGB
  stars and find relatively high median metallicity values
  ($<$[Fe/H]$>_{med} \sim-0.9$ to $-1.0$~dex) that change very little
  with radius over the extent of our survey.  Radial metallicity
  gradients are measured to be $\approx -0.002-0.004$~dex/kpc and the
  fraction of metal-poor stars (defined as [Fe/H]$<-1.0$) is $\approx
  40-50$\% at all radii.  We discuss these findings in the context of
  galaxy formation models for the buildup of gE haloes. 
\end{abstract}



\section{Introduction}
 
Giant elliptical (gE) galaxies contain a large fraction of the stellar
mass in the Universe and it is crucial to understand how they form and
evolve. Important clues come from analyses of their structures,
stellar populations and kinematics and a variety of models have been
invoked in order to reproduce and explain their observed properties.
The classical monolithic collapse models have stars forming during an
early dissipative phase where the gas rapidly flows into the central
galaxy regions \citep[e.g.][]{larson74, carlberg84, arimoto87}.  On
the other hand, in the hierarchical scenario of a $\Lambda$CDM
cosmology, either a few major mergers or several minor mergers and
accretion events are believed to build-up the haloes of the largest
galaxies \citep[e.g.][]{kauffmann93, delucia06}.  Recent work has
favoured a two-stage scenario \citep[e.g.][]{naab07, naab09,
  kaviraj09, oser10, oser12} whereby early in-situ star formation
builds up the central parts of gEs at $z>2$, while later low-mass dry
accretions feed the growth at large radii.  Such models provide a
natural explanation for the observation that some massive ellipticals
at high z are more compact than their low redshift counterparts (e.g.,
\citealt{daddi05, trujillo06,buitrago08}).

Observations confirm the existence of a dominant population of old
stars in the centers of gEs \citep[e.g.][]{thomas05}, but few
constraints exist to date on the properties of stars in regions
beyond the effective radius, $R_{\rm eff}$. The reason for
this is that most studies of gEs have been based on integrated light
analyses, and the faintness of the outer regions has hindered
quantitative study of these parts.  Some work has demonstrated the
existence of faint tidal debris in the outskirts of ellipticals
\citep[e.g.][]{tal09, janowiecki10} while a few recent studies have
started to probe kinematics and chemical composition out to
increasingly large radii \citep[e.g.][]{weijmans09,
  spolaor10a,spolaor10b,coccato10a, coccato10b, greene12,
  labarbera12}.  The general picture emerging is one of older ages and
lower metallicities in the outskirts of gEs, qualitatively consistent
with the predictions of the two-phase model. However, it should be
noted that these studies typically have not probed beyond a few
effective radii.
 
To probe gEs to even more extreme radii, techniques other than
integrated light analysis are required. \citet{tal11} stack SDSS
images of more than 42000 massive red galaxies at $z\sim0.35$ in order
to explore the mean properties to radii of $\sim400$~kpc. They find
the halo colour is bluer than the central regions and remains constant
beyond $\sim3 R_{\rm eff}$. They also find a deviation from a simple
Sersic profile \citep{sersic68} at radii $\gtrsim8 R_{\rm eff}$, in
the sense that there is excess light, and an increasing ellipticity.
While powerful, the stacking technique has the disadvantage that
information regarding individual galaxies is lost and the
interpretation can be complicated if the stacked sample is not truly
homogeneous.

An alternative method is to use resolved stellar populations to probe
the outskirts of gEs.  This technique can only be applied to a small
number of very nearby systems at present, but has the advantage that
age and metallicity can be directly constrained from colour-magnitude
diagram (CMD) morphology, removing several of the uncertainties
inherent in integrated light analyses.  Wide-area resolved star counts
also allow the structure and extent of individual galaxies to be
probed to extremely low effective surface brightness levels,
competitive with those attained through the stacking analysis of many
thousands of systems.  This has already been demonstrated in a number
of recent studies of spiral and dwarf galaxies
\citep[e.g.][]{ferguson02, ibata07, mcconnachie09, barker09, barker11,
  bernard12b} but thus far there has been no wide-field study of the
resolved stellar populations in a gE.

\begin{figure*}
  \centering
  \includegraphics[width=8.5cm]{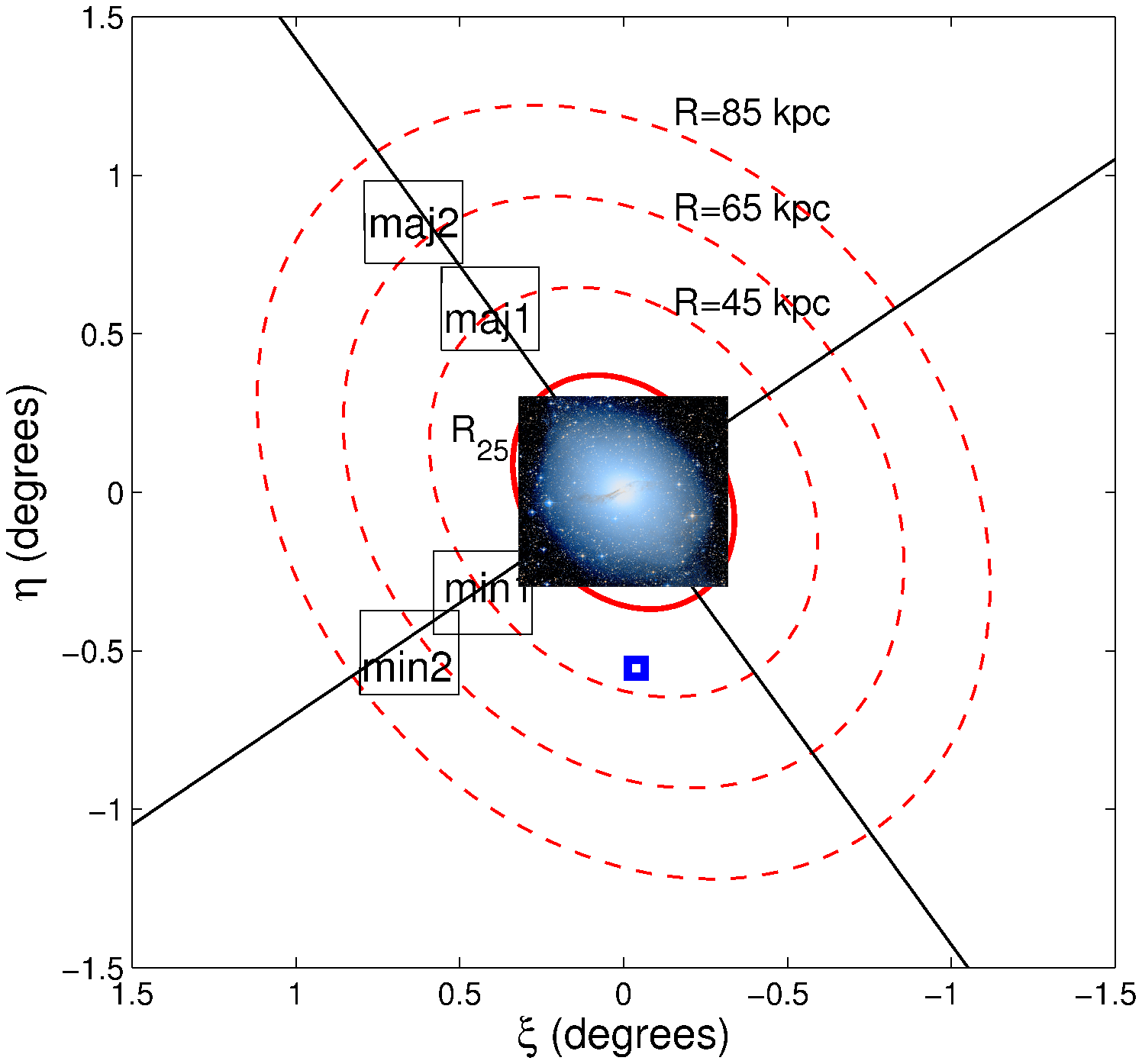}
 \includegraphics[width=8.5cm]{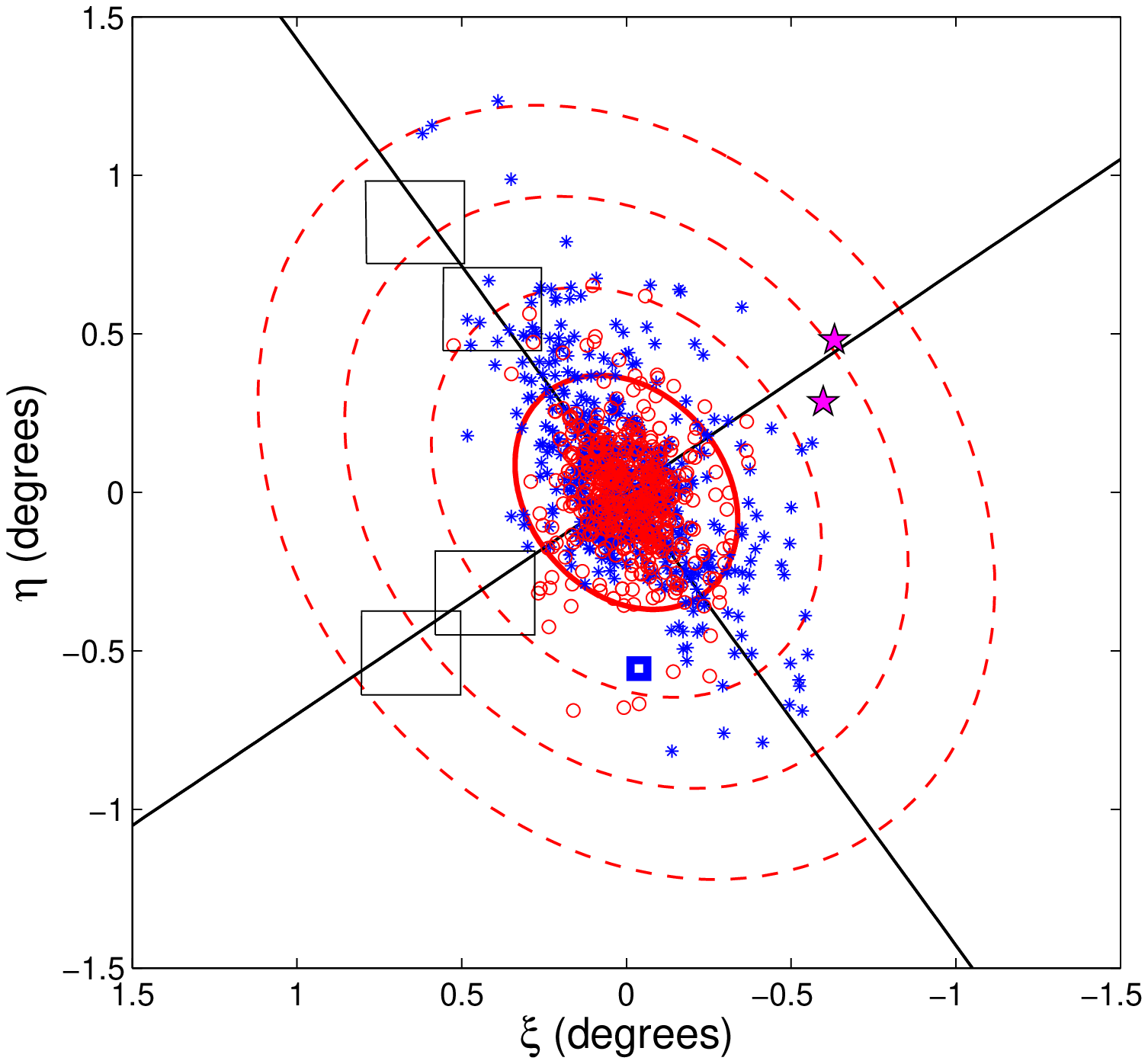}
 \caption{Projected position on the sky in standard coordinates (with
   respect to the center of Cen~A) of the four VIMOS observed fields
   (black rectangles). \emph{Left panel.}  We show a photographic
   plate image of the central regions of Cen~A.  The solid red ellipse
   represents the 25th $B$-band isophote as listed in the RC3
   (R$_{25}=25.7'\sim27.8$ kpc, b/a$=0.77$ and PA$=35^{\circ}$), while
   red dashed ellipses are drawn at projected radii of 45~kpc
   (corresponding to $\sim7.5 R_{\rm eff}$), 65~kpc ($\sim10.5 R_{\rm
     eff}$) and 85~kpc ($\sim14 R_{\rm eff}$). The black solid lines
   represent the major and minor axes. The blue square shows the
   location of the outermost ACS/HST pointing (projected radius of
   $\sim38$~kpc) in the Cen~A halo to date, analysed by
   \citet{rejkuba05,rejkuba11}. \emph{Right panel.} We show the
   positions of confirmed globular clusters (red circles) from the
   \citet{peng04} and \citet{woodley10} catalogs, and confirmed
   planetary nebulae (blue asterisks) from \citet{peng04b}. The latter
   extend out to projected radii of 85~kpc. The two magenta stars
   indicate the positions of two early-type dwarf companions of Cen~A
   (KK197 and KKs55, studied in \citealt{crnojevic10}).}
\label{spatdist}
\end{figure*}

Located at a distance of 3.8 Mpc \citep{harrisg09} and sitting at the
centre of the homonymous group of galaxies, Centaurus~A (Cen~A,
NGC~5128) is the best target for wide-field resolved stellar
populations analysis of a gE from the ground. This peculiar system
hosts an active nucleus and shows striking evidence for having
experienced a recent gas-rich merger event. In particular, the inner
regions ($\lesssim30$ kpc, or within the 25th $B$-band isophote,
R$_{25}$) contain dust lanes, a dense warped gas disk and diffuse
stellar shells and arcs \citep[e.g.][see also Fig.
\ref{spatdist}]{haynes83, malin83, peng02}.  The strong radio emission
from Cen~A partly overlaps with these optical features, but extends
very much further out, covering almost 8 degrees in declination
\citep[for a review, see][]{israel98}.  Nonetheless, with a mass of
$\sim0.5-1\times 10^{12}$~M$_{\sun}$ \citep{woodley07, woodley10c} and
a luminosity of M$_V \sim -21.5$, Cen~A is typical in a global sense
of the field ellipticals seen at low redshift.

Fifteen years ago, the old stellar populations of Cen~A were the first
to be resolved in a galaxy outside the Local Group by using WFPC2
onboard the Hubble Space Telescope (HST) \citep{soria96}.  Since then,
numerous space- and ground-based studies \citep[][]{harris99,harris00,
  harris02, marleau00, rejkuba01, rejkuba03, rejkuba05, rejkuba11,
  crockett12} have surveyed Cen~A in both optical and near-IR (NIR)
bands out to radii of $\sim38$ kpc in projection.  At the distance of
Cen~A, the physical scale is 1 arcmin $\sim1$ kpc and, given that
$R_{\rm eff}\sim330$~arcsec$\sim6.1$~kpc \citep{vandenbergh76}, this
radial extent corresponds to $\sim6.2 R_{\rm eff}$.  Amongst the
results, the predominantly old red giant branch (RGB) population is
found to be moderately metal-rich ([M/H]$\sim-0.65$) with a broad
metallicity spread and no significant metallicity gradient.  In the
deepest HST field studied to date ($\sim38$~kpc), an intermediate-age
population is estimated to account for $\sim20-30\%$ of the
population, and has a derived age of $\sim2-4$ Gyr
\citep[][]{rejkuba11}.  A much younger ($\sim10$ Myr), and more
localized, trace population is found within $\sim8$ kpc, aligned with the
radio jet of Cen~A and possibly triggered by bow shocks propagating
through the interstellar medium \citep{crockett12}.  Although probing
a significant range in radius, these studies are all based on small
field-of-view (FOV, a few arcmin$^2$) imagery, leaving open questions
about how representative they are of the \emph{global} properties of
Cen~A's stellar halo.

There is also good reason to believe that the halo of Cen~A may extend
very far out, well beyond the region where previous resolved stellar
populations studies have probed.  Indeed, globular clusters (GCs) and
planetary nebulae have been discovered out to projected radii of
$\sim85$ kpc \citep[][see also Fig. \ref{spatdist}]{peng04, peng04b,
  woodley07, woodley10, harris12}.  \citet{harris07a} discuss the
halo of another nearby gE, NGC3379, where the outer stellar halo
is found to be predominantly metal-poor at radii $\gtrsim10 R_{\rm
  eff}$. Such a transition from a metal-rich to a
metal-poor halo has not yet been observed in Cen~A suggesting that, if
it is a universal feature of gEs, it may lurk at larger radii.

With this motivation in mind, we have undertaken the first wide-field
survey of the resolved stellar populations in the remote outer halo of
Cen~A.   This work is part of a larger programme we are conducting
to explore the low surface brightness outer regions of galaxies within 
5~Mpc using wide-field imagers on 8-m class telescopes \citep{barker09,
barker11,bernard12b}. In \S \ref{obs} we present our observations and photometric
analysis, and in \S \ref{cmds} we present the resulting CMDs. In \S
\ref{spat} we investigate the structure of Cen~A's outer halo while in
\S \ref{mdfs_sec} we derive metallicity distribution functions (MDFs)
and constrain the radial gradients.  In \S \ref{disc} we discuss our
results, and the summary is presented in \S \ref{concl}.

\section{The Data} \label{obs}
\subsection{Observations} 

Observations were obtained in imaging mode with the VIMOS instrument at
the ESO Very Large Telescope (VLT) on Cerro Paranal, Chile under programme
074.B-0741 (Cycle 74, PI: Ferguson).  VIMOS \citep{lefevre03}
is a visible wide field imager and multi-object spectrograph mounted
on the Nasmyth B platform of the 8.2m Unit Telescope 3 (Melipal).  The
instrument is made up of four identical arms, each with a field of
view of $7\times 8$ arcmin$^2$, separated by $\sim 2$ arcmin.  The
pixel scale is 0.205 arcsec.

\begin{table*}
 \centering
\begin{minipage}{140mm}
  \caption{Observing log for each of the four pointings.}
\label{obslog}
  \begin{tabular}{lcccccccc}
    \hline
    \hline
    Field & $\alpha_{J2000}$ & $\delta_{J2000}$ & R$_{circ}$\footnote{Projected linear distance from Cen~A's center ($\alpha_{J2000}= 13^h 25^m 27.6^s$, $\delta_{J2000}=-43^\circ 01' 08.8''$; taken from the NED database).} & R$_{ell}$\footnote{Projected elliptical radius from Cen~A's center, computed assuming b/a$=0.77$ and PA$=35^{\circ}$ (values from the RC3; \citealt{corwin94}).}  & F & t$_{exp}$&N$_{exp}$&Seeing$_{med}$\\
    &($^h\, ^m\, ^s$)&($^\circ$ ' '')&(arcmin/kpc)&(arcmin/kpc)&&(sec)&&(")\\
    \hline
    {Cen~A-maj1}&13 27 39.2&-42 26 35.2&42.3/45.7&42.3/45.7&$I$&6000&15&0.73\\
    &13 27 40.7&-42 26 30.1&&&$V$&10000&8&0.76\\
    {Cen~A-maj2}&13 28 55.1&-42 09 57.6&63.8/68.9&63.8/68.9&$I$&5200&13&0.68\\
    &13 28 55.0&-42 10 03.7&&&$V$&10000&8&0.58\\
    {Cen~A-min1}&13 27 49.8&-43 20 13.9&32.2/34.8&41.8/45.1&$I$&8000&20&0.57\\
    &13 27 49.8&-43 20 17.2&&&$V$&8750&7&0.66\\
    {Cen~A-min2}&13 29 04.9&-43 31 29.3&49.8/53.8&64.7/69.9&$I$&6800&17&0.68\\
    &13 29 05.0&-43 31 33.6&&&$V$&10000&8&0.60\\
    \hline
    \hline
\end{tabular}
\end{minipage}
\end{table*}

We imaged two fields along each of the major and minor axes, centered
at projected galactocentric elliptical radii of $R \sim45$ and 70~kpc;
these fields all lie well beyond the the known shell system of Cen~A
which is generally confined to $\lesssim$~R$_{25}$ \citep{haynes83,
  malin83, peng02}.  The radii, which refer to the semi-major axis of
the elliptical isophote which intersects a given position, are
calculated assuming an axis ratio of b/a$=0.77$ and a position angle
of $35^{\circ}$ \citep{corwin94}. Throughout this paper, whenever we
refer to a "radius", we mean a projected elliptical radius computed in
this way.  The footprint of each pointing is drawn and labelled in
Fig.  \ref{spatdist} where some fiducial radii are also marked. This
figure also shows the location and size of the deepest extant HST/ACS
pointing which lies at a projected radius of $\sim38$~kpc. Each of our
VIMOS fields was observed with 13 to 20 individual exposures of
$\sim400$ sec in the $I$-band, and 7 to 8 exposures of $\sim1250$ sec
in the $V$-band. Individual exposures were dithered by a few tens of
arcsecs. The seeing was excellent during most of the observing runs,
ranging from 0.45--0.98'' with median values of $\sim0.67$'' for
$I$-band images and $\sim0.65$'' for $V$-band images.  Full details of
the observations are reported in Tab. \ref{obslog}.

The sensitivity of each individual VIMOS detector is lower at the
edges, with additional distortion and vignetting in some cases. Stars
which fall in these regions are thus not recovered well and we exclude
these regions from our analysis. The effective area that was surveyed
by our observations (i.e., after eliminating inter-chip gaps, masking
bad edges, and considering the overlap between the minor axis fields)
is $\sim0.22$ deg$^2$, corresponding to $\sim30\times30$ kpc$^2$.

The data reduction procedure followed a modified version of the
pipeline developed for processing Wide Field Camera data from the
Isaac Newton Telescope (for further details see \citealt{irwin85,
  irwin97, irwin01, irwin04, barker11, bernard12b}). The first stage
includes bias and overscan-correction, plus trimming of each image to
the useful active detector area. Master flats were then created by
stacking a well-exposed dithered set of 21 $V$-band and 9 $I$-band
twilight sky observations. The flat-fielding also corrects for
internal gain differences between the detectors. The dark sky $I$-band
images were combined to form a master fringe frame.  Typical fringe
pattern amplitudes were around $4\%$ of the sky with a range of
spatial scales.  Individual $I$-band science images were corrected for
fringing using a scaled version of the master fringe frame.

Before stacking the science images, object catalogues were generated
for each image to refine the astrometric calibration and to assess the
data quality.  A Zenithal polynomial projection \citep{greisen02} was
found to provide an accurate World Coordinate System (WCS) of the
telescope plus imaging camera.  A third-order polynomial included all
the significant radial field distortions and a further six-parameter
linear model per detector was used to define the remaining astrometric
transformations. The Two Micron All Sky Survey (2MASS) point-source
catalogue \citep{cutri03} was used for the astrometric reference
system.

Common background regions in the overlap area of individual exposures
were used to correct for sky variations during the exposure sequence
and the final stack included seeing weighting, confidence (i.e.,
variance) map weighting and clipping of cosmic rays. First-pass
catalogues using standard aperture photometry techniques
\citep{irwin04} were generated and used to update the astrometric
solution for the stacked images and to provide the aperture
corrections to total flux required for the final photometric
calibration.

In addition to the science targets, a series of Landolt standard
fields \citep{landolt92} were observed throughout the programme.
These were analysed in the same way as the science image and provide
the instrumental to photometric system (Johnson-Cousins) calibration.
The variation in derived magnitude zero-points seen in the standard
field observations was at the level of $\pm2\%$ and we take this to be
indicative of the final accuracy of our photometric calibration.

\subsection{Photometry} 

Although aperture photometry was performed as part of the data
reduction process, we chose to further perform point spread function
(PSF)-fitting photometry on the stacked images so as to more
accurately recover faint stars in crowded regions, obtain additional
morphological information on detected sources and better assess
photometric errors and incompleteness effects.  For this we used the
photometric suite of programs DAOPHOT and ALLFRAME \citep{stetson87,
  stetson94}. For each chip of the stacked images, the PSF was
constructed by using at least 30 bright, non saturated stars, evenly
spread across the image. We derived a list of objects with a $3\sigma$
detection threshold in each filter, and computed the coordinate
transformations between the different filters with the
DAOMATCH/DAOMASTER packages \citep{stetson93}. The list of stars
detected in at least one of the two filters for each chip was used as
the input list for the simultaneous photometry on the stacked images
using ALLFRAME. In the final stellar catalogue, we retain sources that
were detected in both filters with photometric errors smaller than
0.3~mag. Furthermore, we only retain sources that are highly likely to
be stellar by requiring absolute values of the sharpness parameter
$\lesssim2$. At magnitudes fainter than $I\sim23$ , we also require
the $\chi$ parameter to be smaller than 1.4--1.6, depending on chip.

We calibrate the magnitudes obtained with our PSF photometry to the
Johnson-Cousins system $I$- and $V$-bands by selecting bright stars in
common with the aperture photometry catalogue, and then deriving a
constant offset between the two.  The total number of stars remaining
after quality cuts is 104325, of which 37117 are in the inner major
axis field (Cen~A-maj1), 22170 are in the outer major axis field
(Cen~A-maj2), 26922 are in the inner minor axis field (Cen~A-min1),
and 18046 are in the outer minor axis field (Cen~A-min2). There is
partial overlap between one chip in each of the Cen~A-min1 and
Cen~A-min2 pointings. We chose to match the common sources and retain
the photometry from the Cen~A-min1 field, since this is the one with
the longest exposure time and best seeing in the $I$-band, hence
smaller photometric errors.

\subsection{Artificial Star Tests} 

We have performed extensive artificial star tests to quantify the
observational uncertainties in our data.  We have simulated fake stars
for each individual chip in each field with an even spatial
distribution and with colours and magnitudes covering the entire
colour-magnitude space spanned by the observed stars. The total number
of simulated stars per chip is $\sim250000$, divided into $\sim80$
subsets so as to not substantially increase the stellar crowding on
the images.  The artificial stars were added to each chip by using the
measured PSF.  The photometry was then performed in the same way as
for the real data, with the same quality cuts applied. We derive
photometric errors as the difference between the input and the
recovered magnitudes of the fake stars, and assess the incompleteness
of our data by counting the number of stars recovered over the total
number of injected stars.

\begin{table}
 \centering
 \caption{$I_0$-band $50\%$ completeness levels at different colours and elliptical radii.}
\label{complet}
  \begin{tabular}{lccc}
  \hline
  \hline
  $(V-I)_0$ & $0-1$ & $1-2$ & $2-3$\\
\hline
&&R$\le45$~kpc&\\
\hline
{Cen~A-maj}&25.2&24.8&24.2\\
{Cen~A-min}&25.7&25.1&24.3\\
\hline
&&R$>45$~kpc&\\
\hline
{Cen~A-maj}&25.4&25.0&24.4\\
{Cen~A-min}&25.6&25.2&24.4\\
 \hline
 \hline
\end{tabular}
\end{table}

\begin{figure}
  \centering
 \includegraphics[width=8.cm]{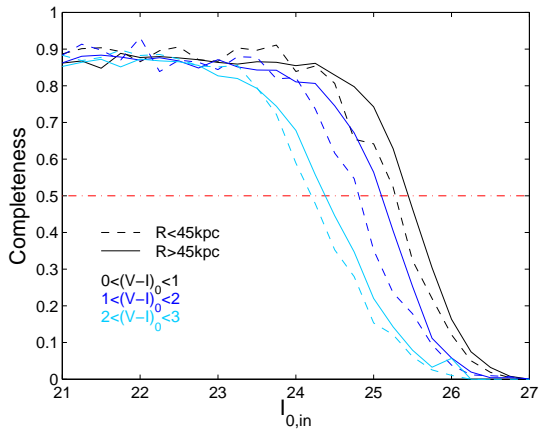}
 \caption{$I_0$-band completeness curves for the four VIMOS fields
   combined (i.e., without distinction between minor and major axes).
   The different curves refer to distinct colour ranges and 
   elliptical radii, as indicated. The dash-dotted red line denotes
   the $50\%$ completeness level.}
\label{complcurves}
\end{figure}

Given that our pointings cover a significant portion of Cen~A's outer
halo, it is important to quantify the photometric uncertainties not
only as a function of magnitude but also as a function of
galactocentric radius.  In Fig. \ref{complcurves} we show the
completeness curves for the entire sample (i.e., without distinction
between major and minor axes) within and beyond $45$~kpc and for
various colour ranges. As expected, the completeness at a fixed
magnitude is somewhat lower at smaller radii because of increased
crowding, and it decreases towards redder colours. The derived
$I_0$-band completeness values for the individual major and minor axes
are given in Tab. \ref{complet}. The radial completeness at a fixed
colour does not vary significantly along the minor axis, partly due to
the fact that the Cen~A-min1 pointing had a longer $I$-band exposure
time than other pointings.  On the other hand, the radial dependence
of the completeness along the major axis is mainly driven by the
intrinsically higher stellar density at smaller radii. Along the major
axis, the typical photometric errors in magnitude are of order
$\sim0.1$ mag at $I_0=22.9$ for radii $\lesssim45$~kpc, and at
$I_0=23.4$ for $R \gtrsim45$~kpc. Along the minor axis, the
photometric errors are $\sim0.1$ mag at a $I_0\sim23.7-23.8$ for all
radii.  This information will be incorporated into our
spatially-resolved analysis of Cen~A's stellar populations whenever
necessary.


\section{Colour-Magnitude Diagrams}  \label{cmds}

\begin{figure*}
  \centering
 \includegraphics[width=8.5cm]{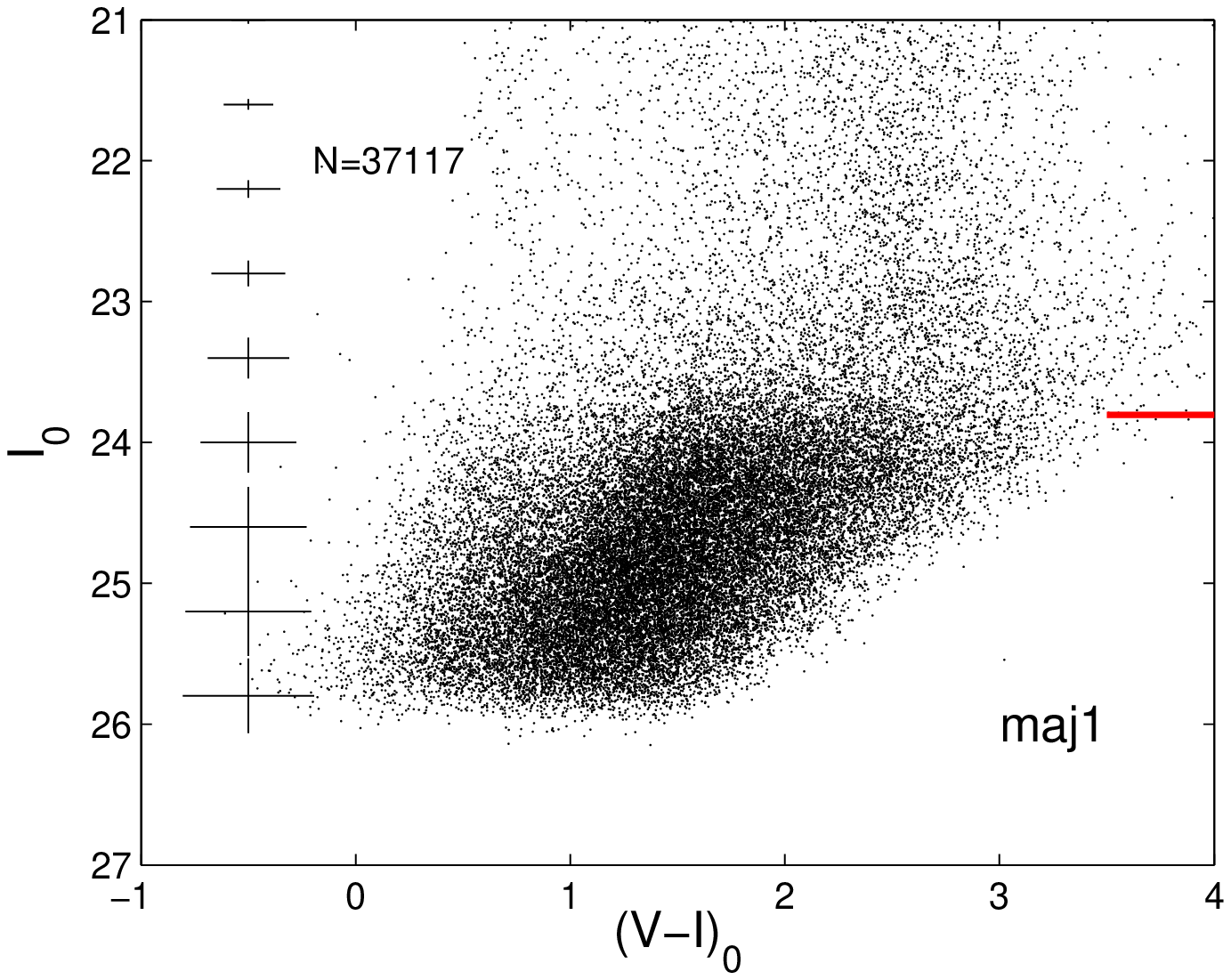}
 \includegraphics[width=8.5cm]{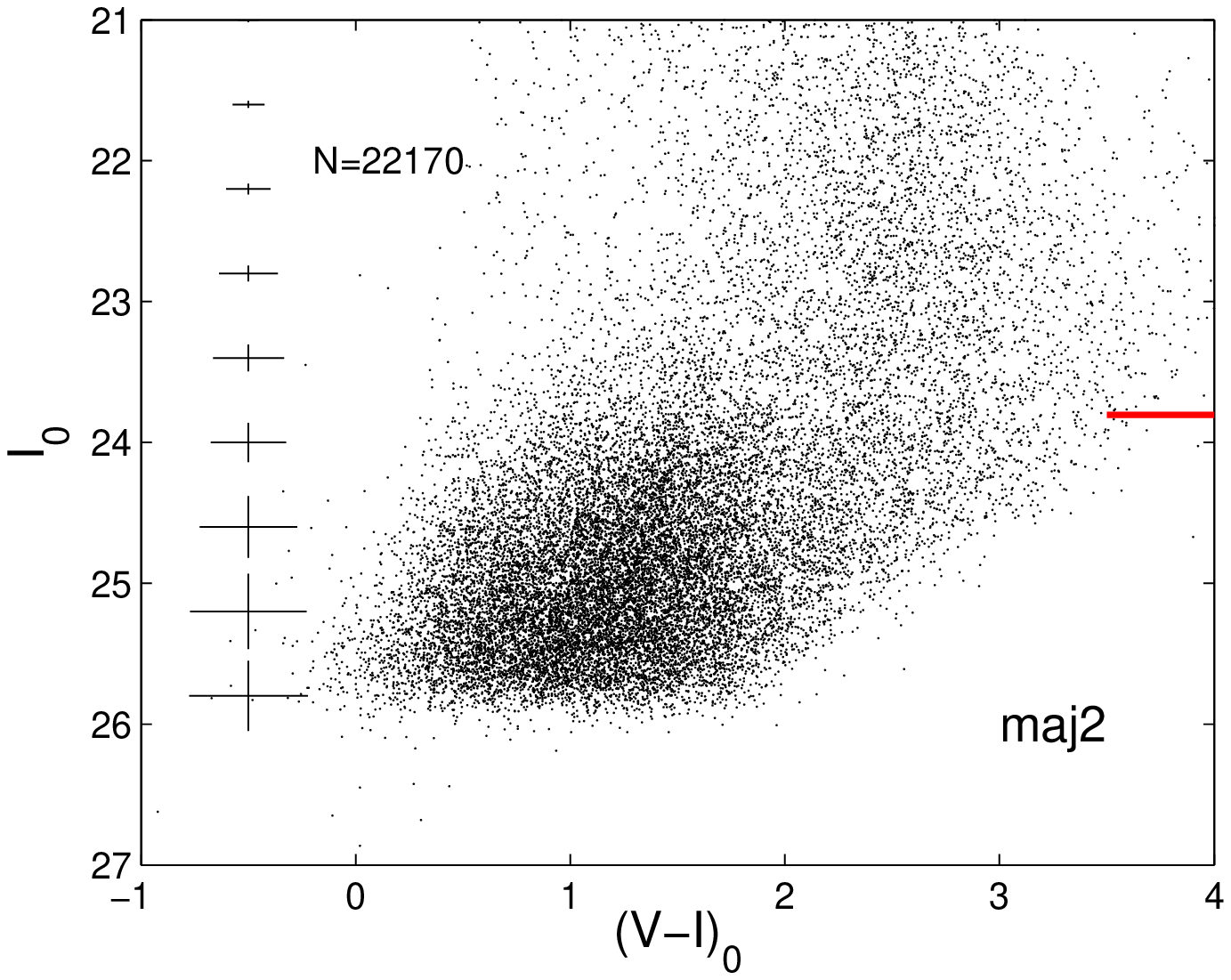}
 \includegraphics[width=8.5cm]{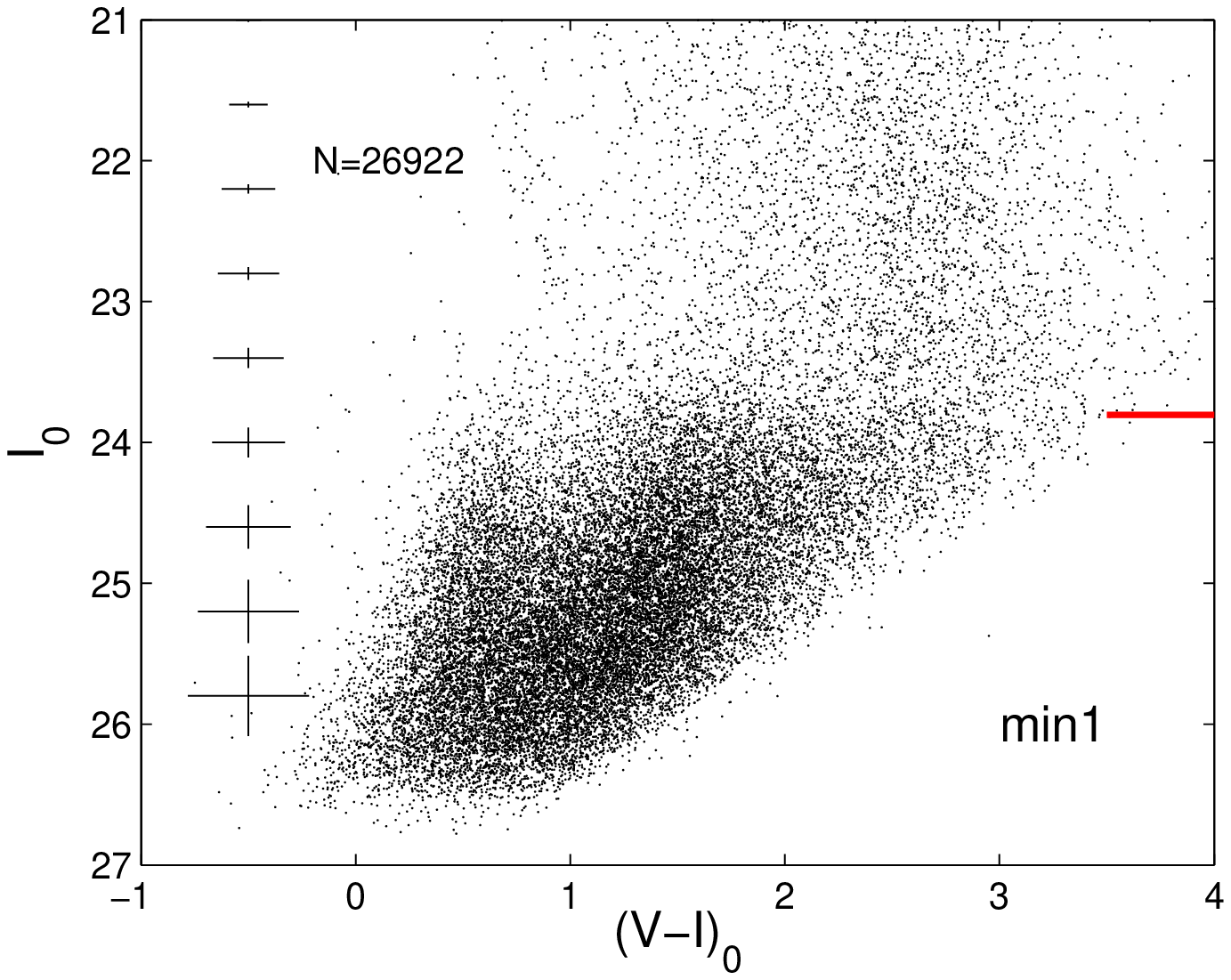}
 \includegraphics[width=8.5cm]{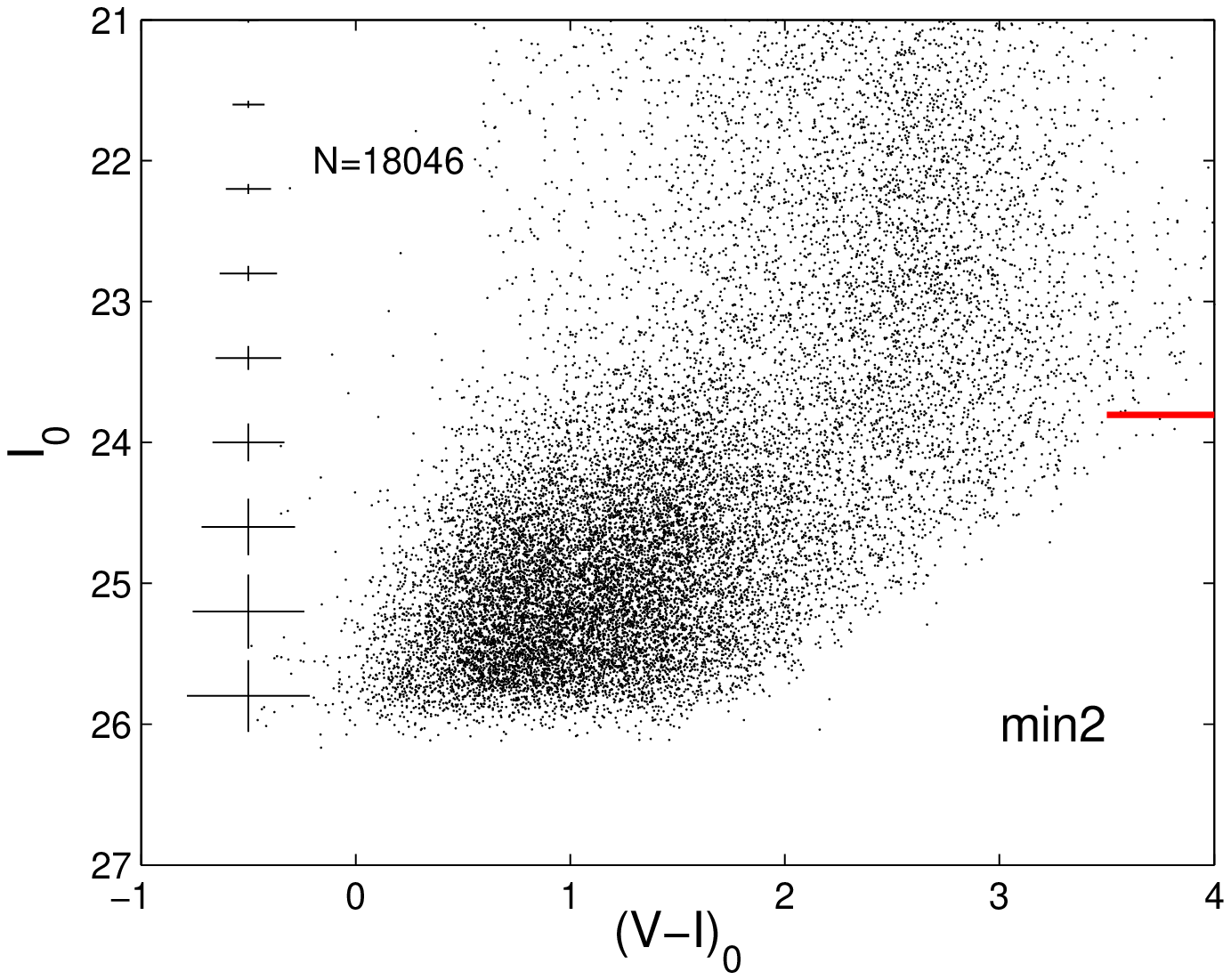}
 \caption{Dereddened CMDs of each individual VIMOS field along the
   major and minor axes (labeled as in Fig. \ref{spatdist}).
   Photometric errorbars are reported, as computed from artificial
   star tests. The red line at $I_{0}=23.81$ aside each CMD indicates
   the average TRGB value derived from the four fields (see text for
   details). }
\label{cmdsfie}
\end{figure*}

The CMDs for each of the four VIMOS pointings are shown in Fig.
\ref{cmdsfie}. We have dereddened the original magnitudes using the
Schlegel extinction maps \citep{schlegel98} and adopting the
\citet{cardelli89} extinction law. In order to account for possible
spatial variations in the reddening (for example, due to the large FOV
and the low Galactic latitude of Cen~A), we have applied extinction
corrections to subfields of 20 arcsec on a side.  On these scales, the
variations are on the order of $\sim0.02$ mag with respect to the mean
extinction values of $A_I\sim0.22$ and $A_V\sim0.38$.  Internal
extinction is expected to be negligible in the gas-deficient outer
regions of Cen~A and we do not take it into account in the subsequent
analysis.

The main feature visible in the CMDs of Fig. \ref{cmdsfie} is a
prominent RGB, a feature indicative of an old population
($\gtrsim1$~Gyr) of evolved cool giants. The upper two magnitudes of
the RGB can be seen in all cases. The width of the RGB is broader than
the photometric errors derived from the artificial star tests implying
an intrinsic spread in colour, likely due to metallicity. The absence
of upper main sequence and blue or red supergiant stars excludes
the presence of stars younger than $\sim1$ Gyr.  In the CMDs, there
are also some stars brighter than the tip of the RGB (TRGB) which 
are candidate intermediate-age ($\sim1-8$ Gyr) luminous
asymptotic-giant branch (AGB) stars.  However, as discussed below,
this region of the CMD is also heavily contaminated by foreground
Galactic stars. Similarily, we will also show that objects in the
colour range $0.3\lesssim (V-I)_0 \lesssim1$ are for the most part
unresolved background galaxies.

In the top panel of Fig. \ref{cmdsfore} we show the combined CMD for
all stars in our photometric catalogue.  We overlay $\alpha$-enhanced
([$\alpha$/Fe] $=+0.2$)\footnote{The choice of the
  $\alpha$-enhancement is justified by the mean value found for Cen~A
  GCs, [$\alpha$/Fe]$\sim+0.14$ \citep{woodley09}.} stellar isochrones
from the Dartmouth group \citep[][]{dotter08} shifted to the distance
of Cen~A with a fixed old age (12 Gyr) and metallicity varying from
[Fe/H] $=-2.5$ to $-0.3$.  For our choice of $\alpha$-enhancement,
these values translate into a range of [M/H] from $\sim-2.4$ to
$-0.15$~dex using the prescriptions of \citet{salaris93}. This set of
theoretical models has been shown to give a particularly good fit to
old and intermediate-age star cluster fiducials
\citep[e.g.][]{glatt08b}. We also show the AGB phase of the Padova
stellar isochrones \citep{girardi10}. We choose isochrones with a
metallicity of $Z=0.004$ (corresponding to [M/H]$\sim-0.7$) and ages
of 2 and 4 Gyr (following the age estimates of \citealt{rejkuba11} for
fields at smaller radii).

Fig. \ref{cmdsfore} also shows the selection boxes used to define the
locations of the RGB and AGB candidates that we will subsequently
analyse. For the RGB box, the faint limit is set by the $\sim50\%$
completeness limit for radii $\gtrsim45$~kpc while the bright limit is
determined by considering the location of the TRGB.  The blue end of
the RGB box is defined by the most metal-poor isochrone and excludes
the bulk of unresolved contaminants. For the AGB box, we start
$\sim0.1$ mag above the TRGB in order to avoid RGB stars scattered up
by photometric errors and allow for a luminosity width of 1.5
magnitudes.  The colours range from the bluest edge of the RGB to
$(V-I)_0 \sim2.8$, thus not encompassing the reddest stars in the CMD.
Stars with colours redder than this might be dust-enshrouded
carbon-rich AGB stars, however only a handful of these are likely to
be present and their numbers will be dwarfed by the foreground
contaminants which dominate at these colours.

\subsection{The Distance to Cen~A}

The TRGB has a fixed $I$-band absolute magnitude of $M_I\sim-4.05$ for
predominantly metal-poor old populations \citep[e.g.,][]{rizzi07} and
can therefore be used as a powerful distance indicator.
\citet{harrisg09} reviews Cen~A's distance measurements to date, as
derived with different indicators. Based on their previous HST
studies, they report a mean $I_{TRGB}=24.10\pm0.1$ which, when
combined with their assumed $A_I=0.22\pm0.02$, yields
$I_{0,TRGB}=23.88\pm0.1$. We recompute the TRGB values for each of our
observed fields using a Sobel edge-detection filter \citep{lee93} and
average them to find $I_{0,TRGB}=23.81\pm0.35$, in excellent agreement
with previous estimates.  The significant uncertainty on our value
comes from a combination of zero-point and extinction uncertainties,
and our chosen binwidth (which depends on the photometric error at the
TRGB value).  The precise absolute magnitude of the TRGB has a
colour/metallicity dependence \citep[e.g.][]{madore09}, becoming
fainter for metal-rich populations ([Fe/H]$\gtrsim-0.5$).  The value
reported by \citet{harrisg09} is computed considering only the
metal-poor side of the TRGB. Our data do not reach the reddest
magnitudes of Cen~A's RGB ($(V-I)_0\gtrsim3$) because of
incompleteness, so we too are sampling a region where the TRGB
magnitude is mostly constant. To be conservative, we used stars with
$(V-I)_0\lesssim2.5$. Finally, we exclude stars bluer than
$(V-I)_0\sim1$ for the TRGB computation, as these are largely
contaminated by unresolved background galaxies. Throughout the paper
we will use the TRGB value derived in this study.

\subsection{Background and Foreground Contaminants} \label{contam}

Even under excellent seeing conditions, a substantial number of
high-redshift background galaxies will appear as unresolved point
sources in ground-based data. Moreover, the low Galactic latitude of
Cen~A ($b\sim20^o$) means that a non-negligible number of Milky Way
dwarf stars will be projected along our sight lines.  Minimizing this
contamination, and ultimately correcting for it, are necessary before
we can properly interpret our data.

\begin{figure}
  \centering
 \includegraphics[width=8.cm]{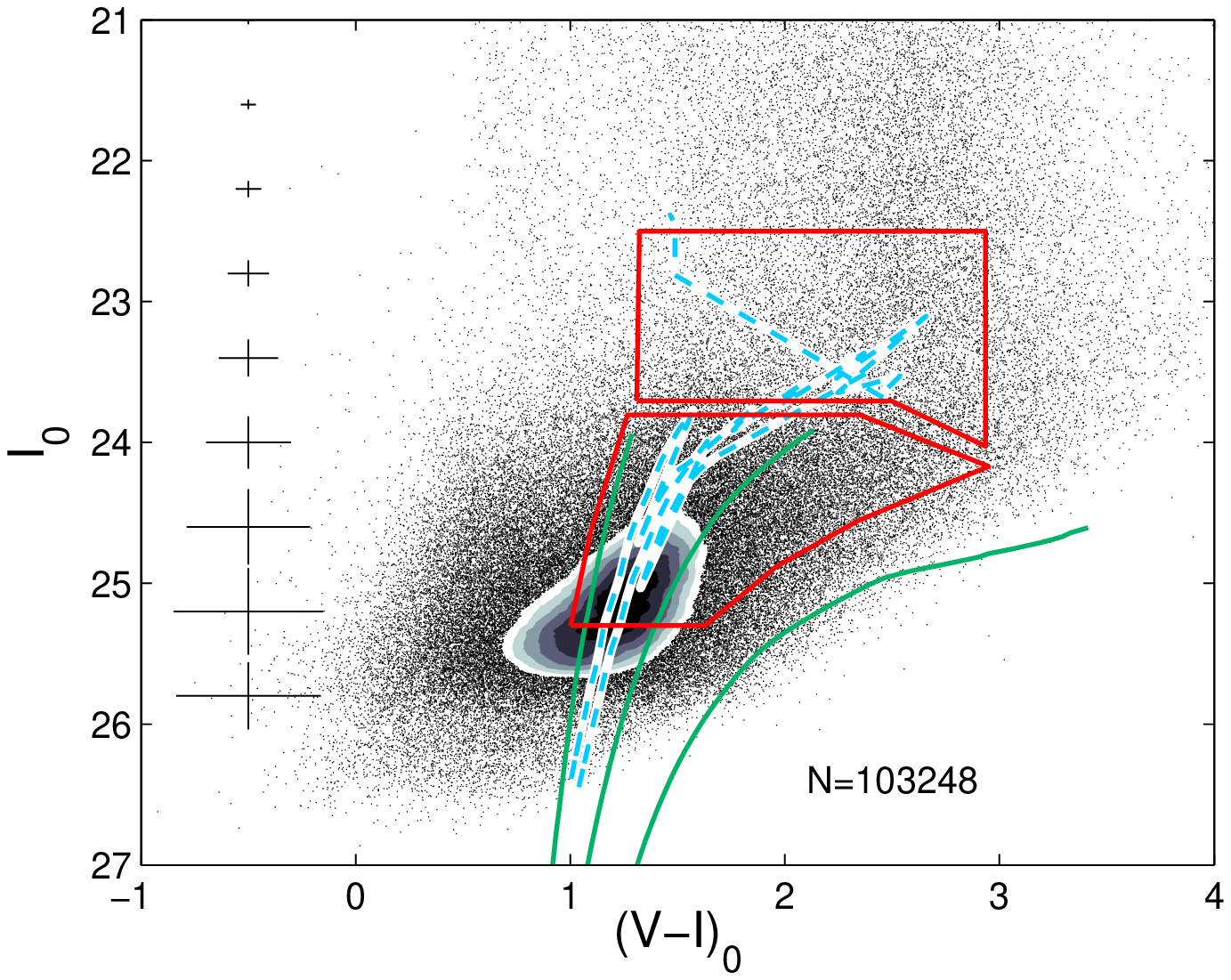}
 \includegraphics[width=8.cm]{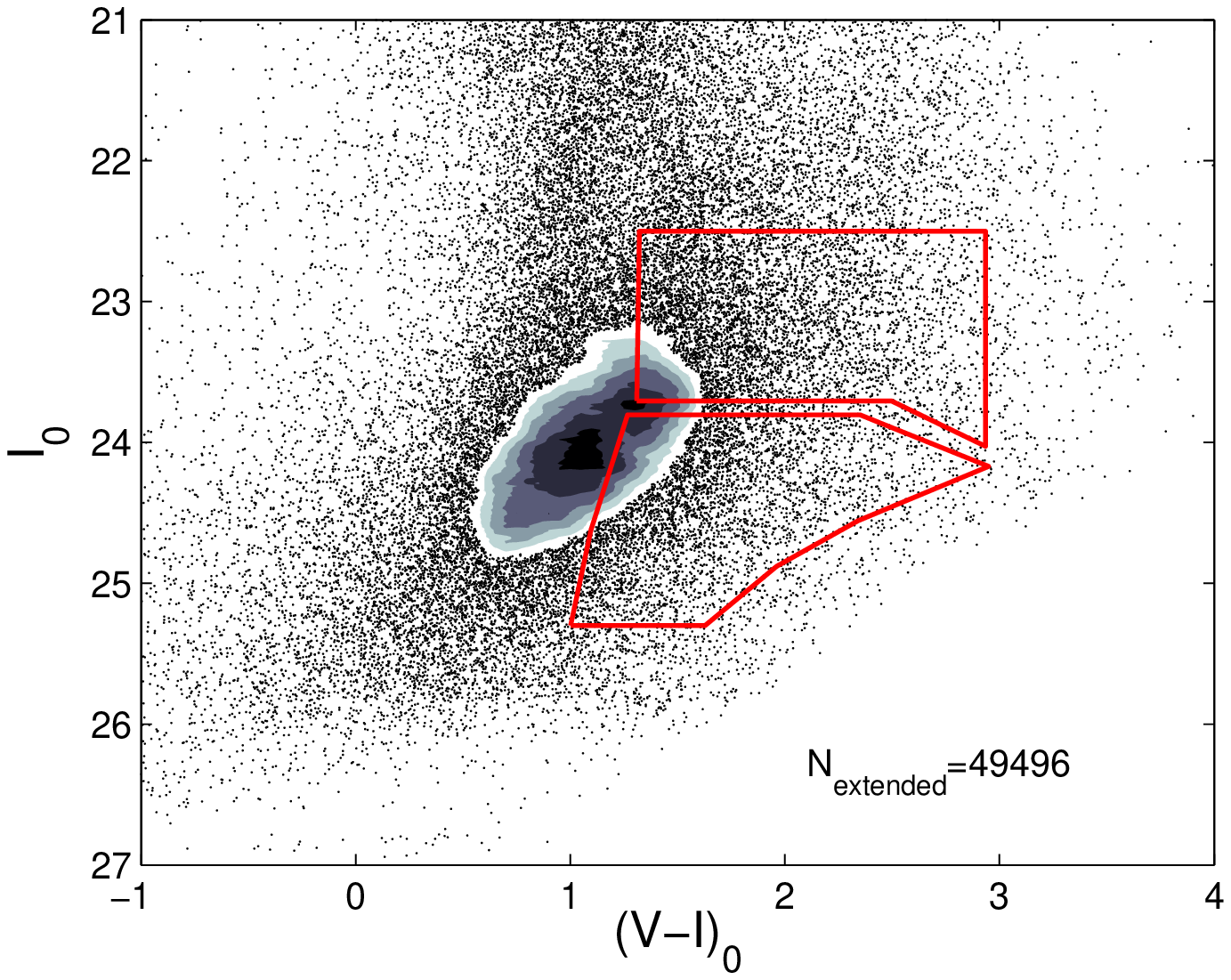}
 \includegraphics[width=8.cm]{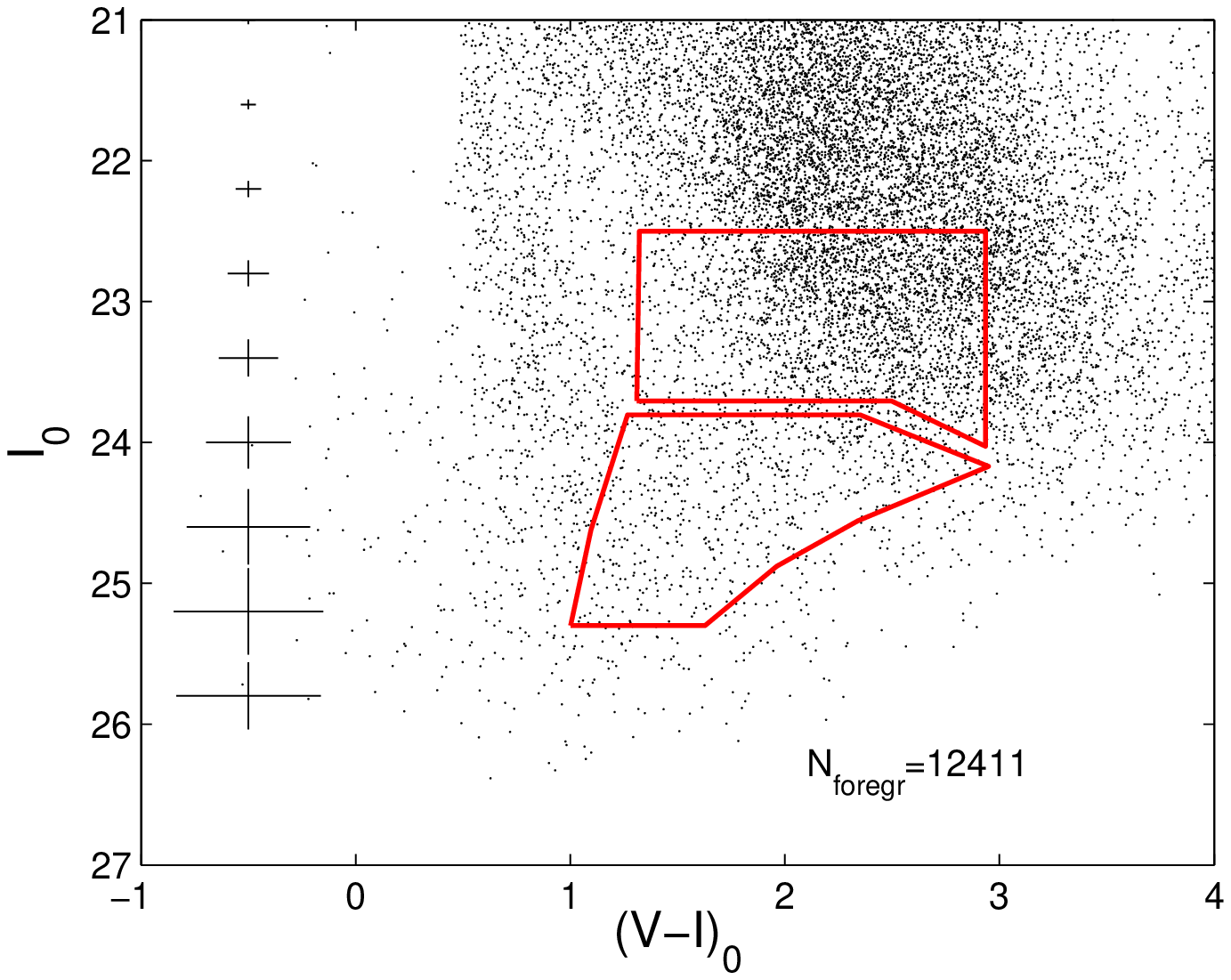}
 \caption{\emph{Top panel}. Dereddened CMD for the combined four VIMOS
   pointings. Individual points are replaced by a Hess (density)
   diagram where the density is highest. We show the boxes used to select the RGB and
   AGB subsamples (red boxes). We overlay Dartmouth stellar isochrones
   for the RGB stars (green lines), with a fixed age (12 Gyr) and
   varying metallicities ([Fe/H] $=-2.5$, $-0.9$ and $-0.3$). We also
   plot the post He-burning AGB phase of the Padova isochrones
   \citep[cyan dashed lines, ][]{girardi10}. The latter have ages of 2
   and 4 Gyr and a metallicity of $Z=0.004$ (see text for details).
   \emph{Central panel}. CMD of all the extended objects rejected by
   quality cuts in our PSF photometry. A Hess diagram indicates the
   highest density region.  \emph{Bottom panel}. CMD of Galactic
   foreground stars as predicted by the Besan\c{c}on models
   \citep[][see text for details]{robin03}.  The foreground
   simulations have been convolved with photometric errors and the
   observational incompleteness effects have been taken into account
   (see text for details). In the central and bottom panels we also
   overplot our RGB and AGB selection boxes to ease the comparison
   with the observed CMDs. }
\label{cmdsfore}
\end{figure}

The min1 field was observed with the best combination of depth and
seeing of all our fields.  Fig. \ref{cmdsfie} shows that this CMD has two
distinct almost parallel sequences, the particular definition of which is due to the
smaller photometric errors.  The blue sequence lies blueward of plausible 
RGB tracks (see top panel of Fig. \ref{cmdsfore}), but redward of where 
we would expect to see blue plume stars ($V-I\sim 0$).  Similarly, red and blue 
supergiants at the distance of Cen~A would be expected at slightly redder and bluer 
colours respectively but at brighter magnitudes than the sequence seen (see Fig. 9 of 
\citealt{barker11}). As in previous work, we therefore identify the blue sequence as resulting 
from unresolved background galaxies \citep{barker09, barker11, bernard12b}.  The fact that 
the surface density of objects in this region of the CMD does not show any trend with 
galactocentric radius adds further weight to it being a population unassociated with Cen A.
To further support our interpretation, we additionally show in the middle panel of Fig.
\ref{cmdsfore} the CMD for all the sources rejected as stars by quality cuts in our PSF 
photometry.  These sources will largely be resolved galaxies.  As can be seen, the dominant 
sequence here does indeed lie blueward of the main RGB sequence.  We have used this fact
to guide our choice of RGB selection box boundaries so as to limit the contamination by these 
non-stellar sources.   While it is clearly impossible to cleanly  separate stars from unresolved 
galaxies through color-magnitude cuts alone, our  conservative approach is designed to 
optimise the fidelity of our RGB catalogue, if not its completeness.

To quantify the contamination from the Milky Way foreground, we have
simulated the CMD of the Galactic population in this direction using
the Besan\c{c}on model \citep{robin03}.  As inputs, we provide the
Galactic coordinates and approximate areas of each of the four
observed VIMOS fields. We apply the photometric errors derived from
our artificial star tests to the simulated stars, and account for
incompleteness by randomly extracting a subsample reflecting the
derived recovery level in each magnitude bin. The bottom panel of
Fig. \ref{cmdsfore} shows the simulated Galactic CMD for the four
pointings combined.  The densest sequences here are clearly
recognisable in the observed CMDs and only minimally contaminate our
RGB selection box.  However, the Galactic populations severely
contaminate the AGB selection box, where they contribute up to
$\sim50\%$ of the sources along the minor axis.  Overall, the
predicted number of foreground stars accounts for $\sim10-15\%$ of the
total number of stars observed.

Consideration of the magnitudes and colours of the background and
foreground contaminants therefore leads us to conclude that our
analysis of Cen~A's remote RGB population will not be strongly
affected by their presence.  In the subsequent sections, we will use
two methods to estimate the actual contaminant source density in our
RGB selection box. In the first instance, we will use knowledge of the
surface density of objects rejected as stars (i.e. that are resolved)
that fall within the RGB selection box to provide a lower limit on the
contaminant level.  
We will also use the surface density of RGB stars in the $75<R<85$~kpc radial
bin along the minor axis as an alternative estimate of the contaminant
level. This region has the lowest point source density of all those
studied in our survey.  Since the number of genuine Cen~A RGB stars
which reside in this area is unknown, it strictly provides an upper
limit on the contamination.  The situation is more difficult for the
AGB selection box and we resort to using model predictions for the
surface density of Galactic stars to assess the contaminant level
here.


\section{Halo Extent and Structure} \label{spat}

\subsection{Colour-Magnitude Diagrams as a Function of Radius} \label{cmds_rad}

\begin{figure*}
  \centering
 \includegraphics[width=18cm]{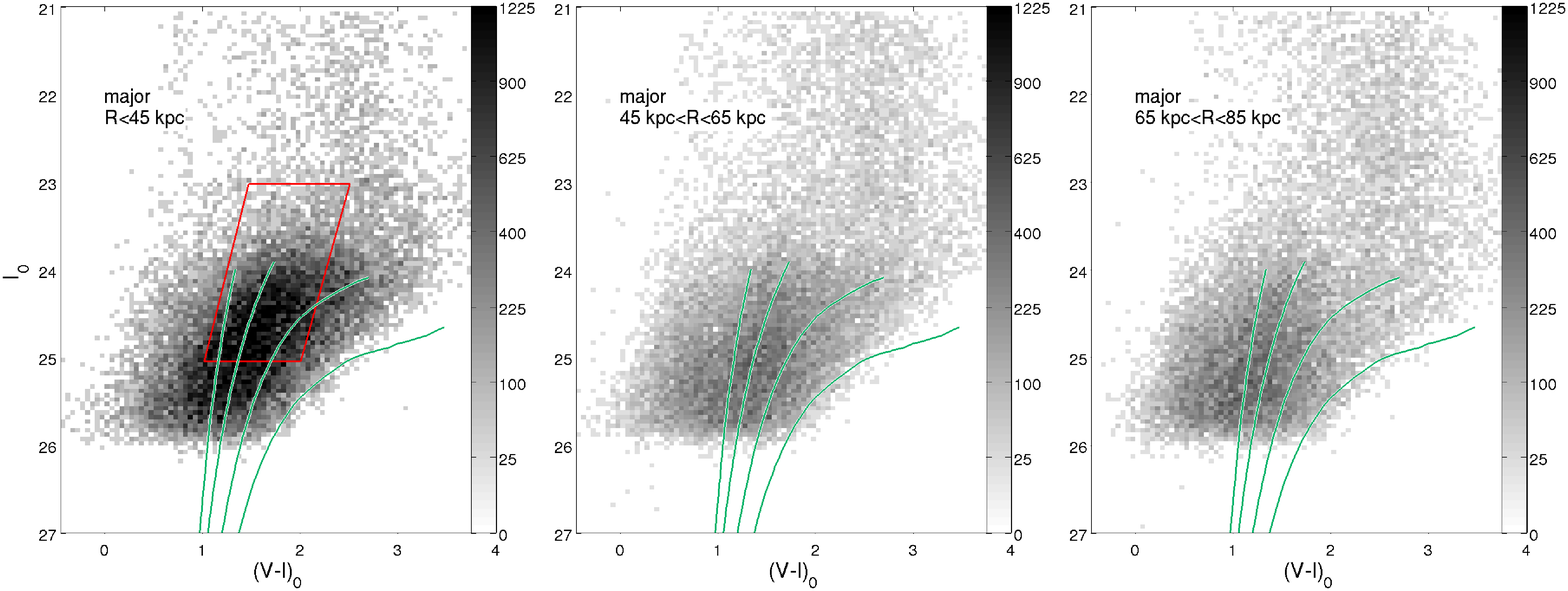}
 \includegraphics[width=18cm]{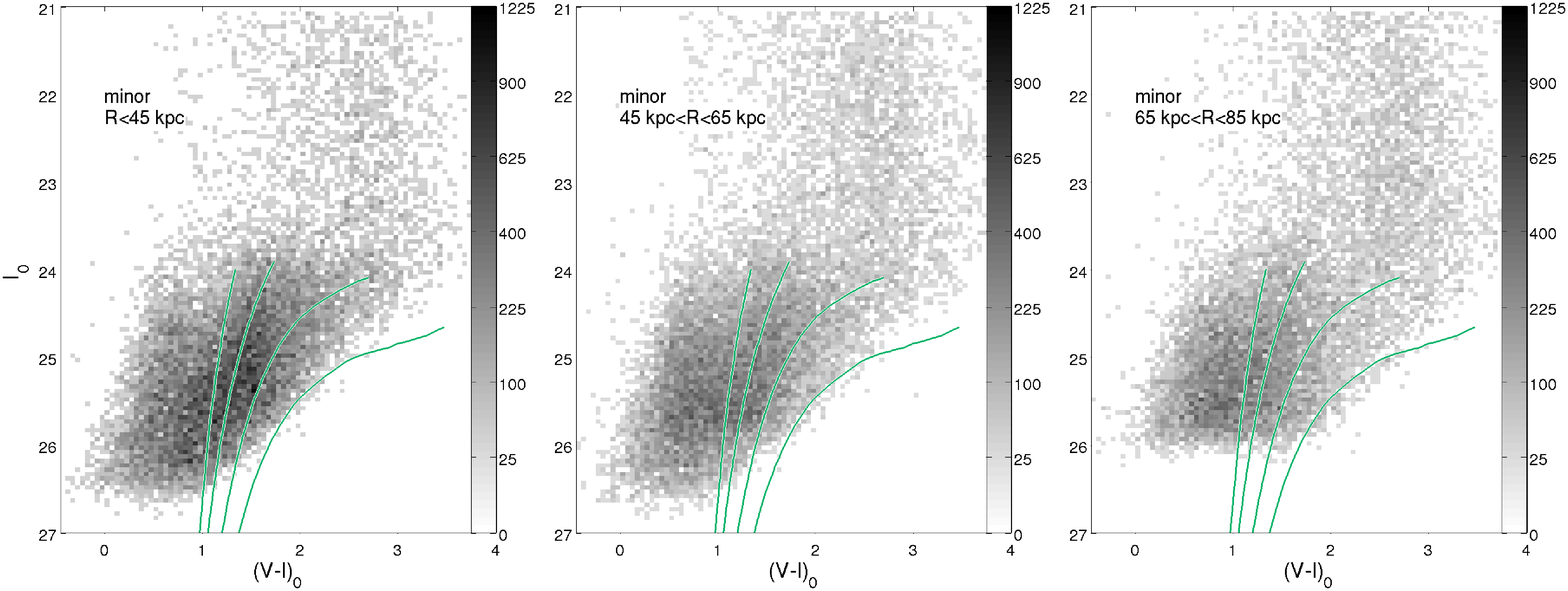}
 \caption{Dereddened Hess (density) diagrams as a function of
    elliptical radius, for both major (\emph{upper panels})
   and minor (\emph{lower panels}) axes. We subdivide the stellar
   samples into three elliptical bins per axis (see also Fig.
   \ref{spatdist}). The colourbars (on a square root scale) indicate
   the number of stars per unit area and per $0.05 \times 0.05$
   mag$^2$. We overlay Dartmouth stellar isochrones with a fixed age
   (12 Gyr) and varying metallicities ([Fe/H] $=-2.5$, $-1.3$, $-0.7$
   and $-0.3$ dex). Note the different stellar densities among the
   radial bins. Finally, in the upper left panel we also 
   show the selection box used to derive LFs (see Sect. \ref{lfs_sec}).}
\label{cmdsell}
\end{figure*}

\begin{figure*}
  \centering
 \includegraphics[width=8.cm]{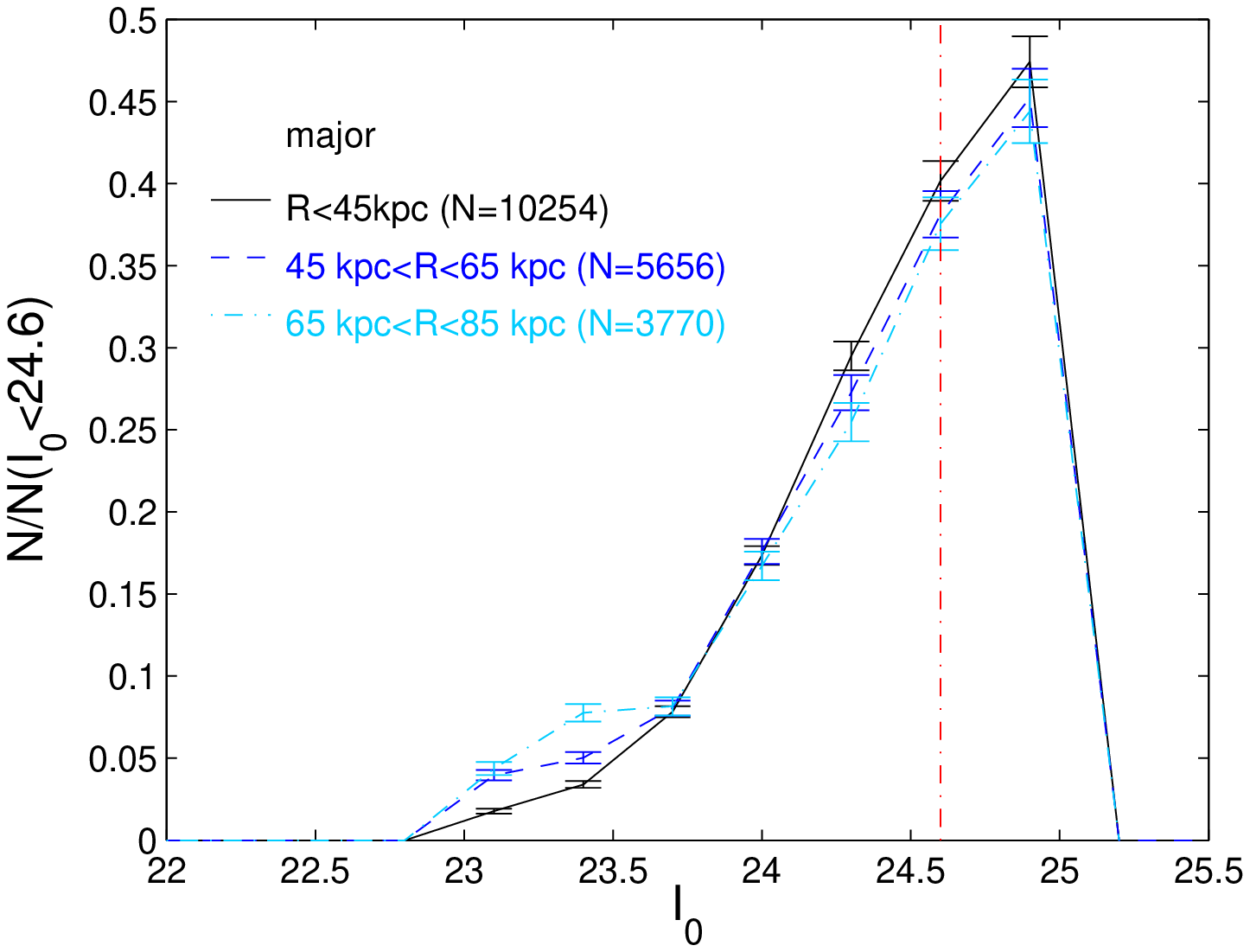}
 \includegraphics[width=8.cm]{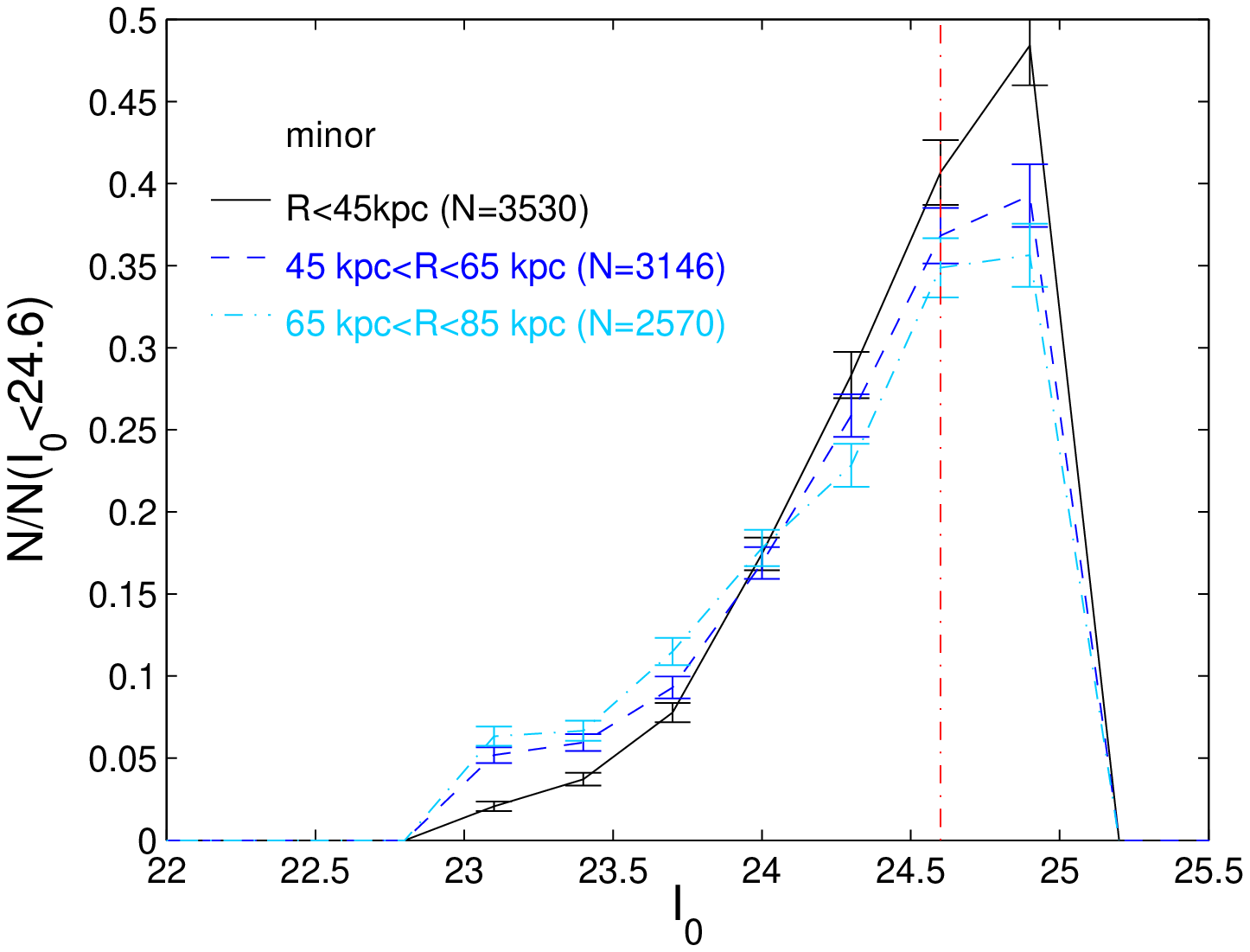}
 \caption{Luminosity functions at different elliptical radii
   as indicated, for both major (\emph{left panel}) and minor
   (\emph{right panel}) axes. The chosen elliptical bins are the same
   as in Fig. \ref{cmdsell}. The LFs are derived from stars in a box
   with $23< I_0 < 25$ and $1 < (V-I)_0 < 2$ (bottom edge) and $1.4 <
   (V-I)_0 < 2.5$ (top edge). They are normalized to the number of
   counts for $I_0\le24.6$ (vertical red dash-dotted lines, i.e., the
   $50\%$ completeness limit for the $R<45$~kpc subsample along the
   major axis). For each LF we report the number counts for
   $I_0\le24.6$. The LFs are corrected for incompleteness, and
   simulated foreground counts have been subtracted for each magnitude
   bin. The errorbars are poissonian.}
\label{lfs}
\end{figure*}

In Fig. \ref{cmdsfie} we show the CMDs for each of our individual
pointings; we now consider the CMDs as a function of radius in order
to search for radial variations.  Fig. \ref{cmdsell} shows Hess
diagrams for three radial bins along each of the major and minor axes
with Dartmouth stellar isochrones overlaid.  Stellar density decreases
with radius along both axes, with the highest density measured
anywhere occurring at $R<45$~kpc along the major axis.  A clear RGB
sequence can be seen at all distances along the major axis, testifying
to the existence of a halo population at least out to $\sim 85$~kpc in
this direction.  On the other hand, it is difficult to discern whether
an RGB sequence is present in the outermost bin ($R>65$~kpc) along the
minor axis.  This asymmetry in the density of stars along the major
and minor axes of the outermost elliptical bin suggests that the outer
halo is either more flattened than the inner isophotes would suggest
or that it is spatially inhomogeneous on the scales probed by VIMOS.
The red extension of the RGB appears to become bluer with increasing
radius indicating a decrease in the number of metal-rich stars.
However, the colour separation between isochrones does not scale
linearly with metallicity and we will subsequently see that this
behaviour does not translate into a strong metallicity gradient (see
Sect. \ref{mdfs_sec}).

\subsection{Luminosity Functions as a Function of Radius} \label{lfs_sec}

In addition to CMDs, we also examine constraints on halo extent that
come from consideration of the luminosity functions (LFs). We
construct the LFs by selecting stars in a colour-magnitude box that
minimizes the background/foreground contamination as much as possible.
Stars are chosen in the magnitude range $23< I_0 < 25$ and in a
diagonal colour range with the bottom edge comprising $1 < (V-I)_0 <
2$ and top edge comprising $1.4 < (V-I)_0 < 2.5$ (see Fig.
  \ref{cmdsell}). We choose magnitude bins of $\sim0.3$ mag, the
largest photometric error within the adopted colour-magnitude range
for the innermost radial bins, and derive histograms in the same
radial bins as before. The LFs are corrected for incompleteness as a
function of magnitude and radius, and the expected foreground counts,
as simulated by the Besan\c{c}on models, are subtracted off.  Finally,
in order to facilitate comparison, the LFs in each bin are normalized
to the total number of stars brighter than $I_0=24.6$ (corresponding
to the $50\%$ completeness limit for the most crowded bin, i.e.,
$R<45$~kpc along the major axis).

\begin{figure}
  \centering
 \includegraphics[width=8.cm]{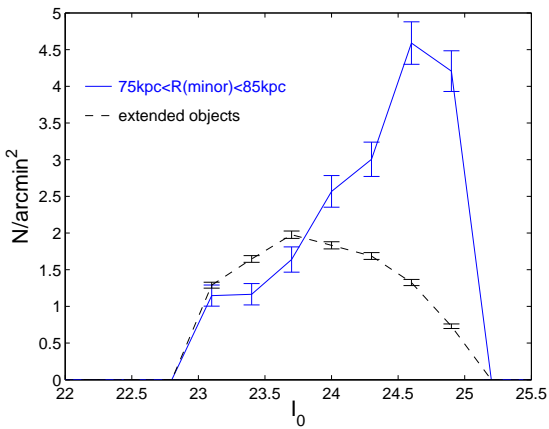}
 \caption{Luminosity functions for the $75<R<85$~kpc field along the
   minor axis (blue line) and for the extended sources rejected by our
   photometric cuts (black dashed line). The number counts are per
   unit area, and both curves are not corrected for incompleteness.}
\label{bck}
\end{figure}

Fig. \ref{lfs} shows the LFs in radial bins along the major and minor
axes.  Poissonian errorbars are shown, which include the uncertainty
in the completeness corrections and in the Besan\c{c}on model counts.
Although the LFs look broadly similar, the outer bins show an excess
of sources brighter than the TRGB ($I_{0,TRGB}=23.81\pm0.35$) with
respect to the innermost bin.  Additionally, the LFs become
progressively flatter at faint magnitudes in the outer radial bins
with the variation being more pronounced along the minor axis.  A
Kolmogorov-Smirnov (KS) test rejects the null hypothesis that the LFs
at radii beyond 45~kpc come from the same underlying population
at a $99 \%$ significance level.

Although these differences could result from genuine radial variations
in the halo populations, it is also plausible that they result from
varying relative contributions of Cen~A stars and contaminants.
Indeed, in the outer halo the latter will contribute an increasing
fraction of the overall source counts and will thus have a more
profound impact on the LF shape.  In Fig. \ref{bck} we plot the LF per
unit area of sources rejected as stars by our quality cuts, derived
from the same selection box as above.  If unresolved background
contaminants follow the same luminosity distribution as resolved
galaxies, then we can assume this is how these objects will contribute
to the overall LF.  We also show in Fig. \ref{bck} the LF per unit
area of point sources in the $75<R<85$~kpc radial bin along the minor
axis, which has not been corrected for incompleteness or foreground
subtracted (we use the bin $75<R<85$~kpc rather than $65<R<85$~kpc in
order to avoid as much as possible the presence of Cen~A halo stars).
Despite the fact that the true fraction of contaminants remains
unknown, this comparison illustrates the excellent agreement between
the two estimates of background LF shapes at magnitudes brighter than
the TRGB. This is reassuring as it indicates that the excess of bright
stars with respect to the innermost radial bins is very likely due to
an increasing fraction of contaminants at large radii. At fainter
magnitudes, the LF of resolved sources turns over with respect to that
of point sources in the $75<R<85$~kpc minor axis bin, consistent with
the progressive flattening seen in the outer LFs in Fig. \ref{lfs}.
Clearly some of this behaviour will be a result of incompleteness in
the resolved source counts but we have no way to correct for this.
Without an accurate measurement of the shape and normalisation of the
contaminant LF, it is impossible to quantify what fraction of the
faint sources in the $75<R<85$~kpc bin LF are genuine Cen~A halo stars
however it is likely that this number is non-negligible.  Indeed, we
have derived TRGB values for each radial bin along both the major and
minor axes and find all of them to be consistent with the value
derived from the total stellar sample (see Sect. \ref{cmds}), within
the uncertainties; this supports the idea that a significant RGB
population is present in all our fields.

\subsection{Radial Density Distribution} \label{densprof_sec}

\begin{figure*}
  \centering
 \includegraphics[width=8.cm]{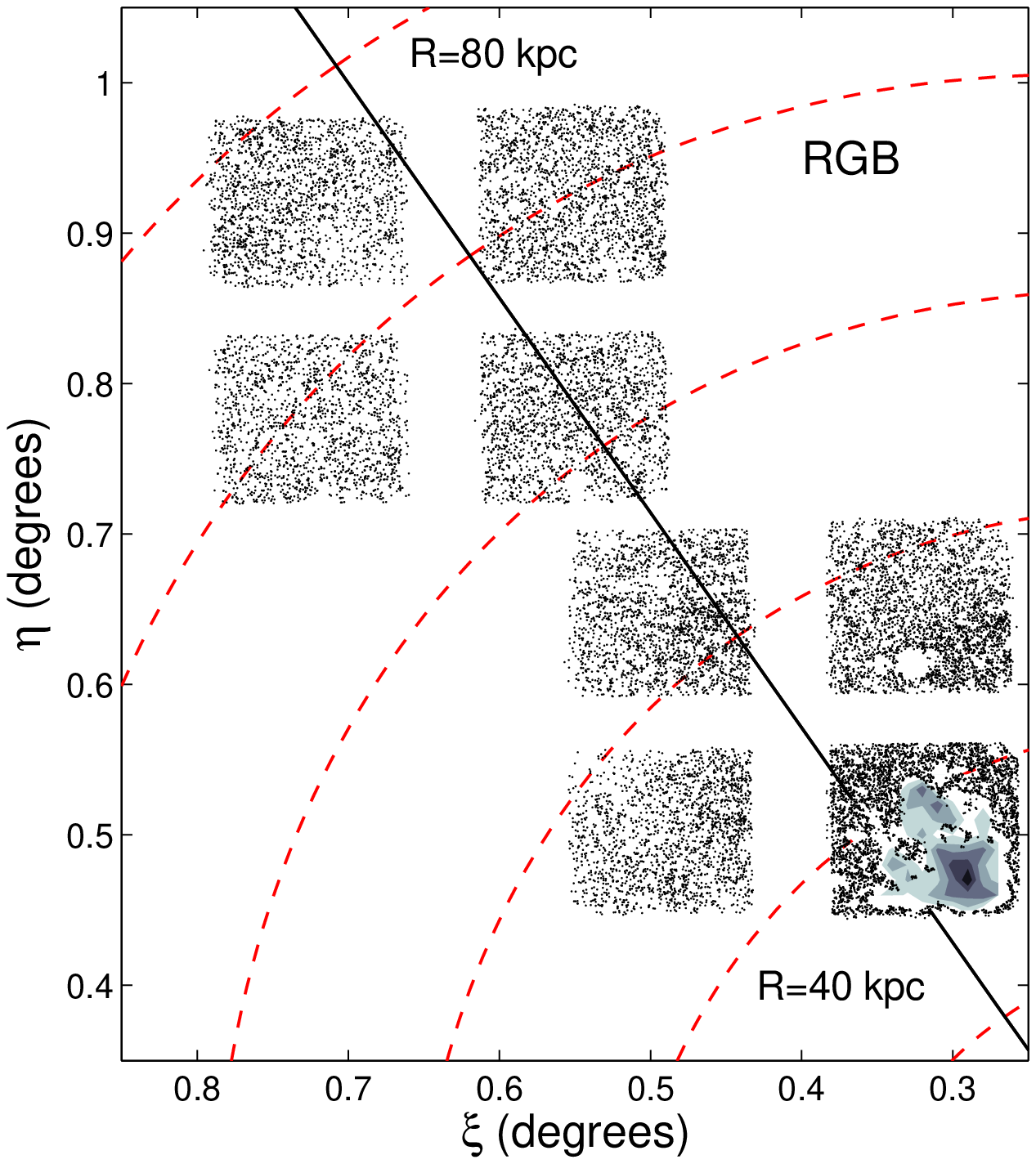}
 \includegraphics[width=8.cm]{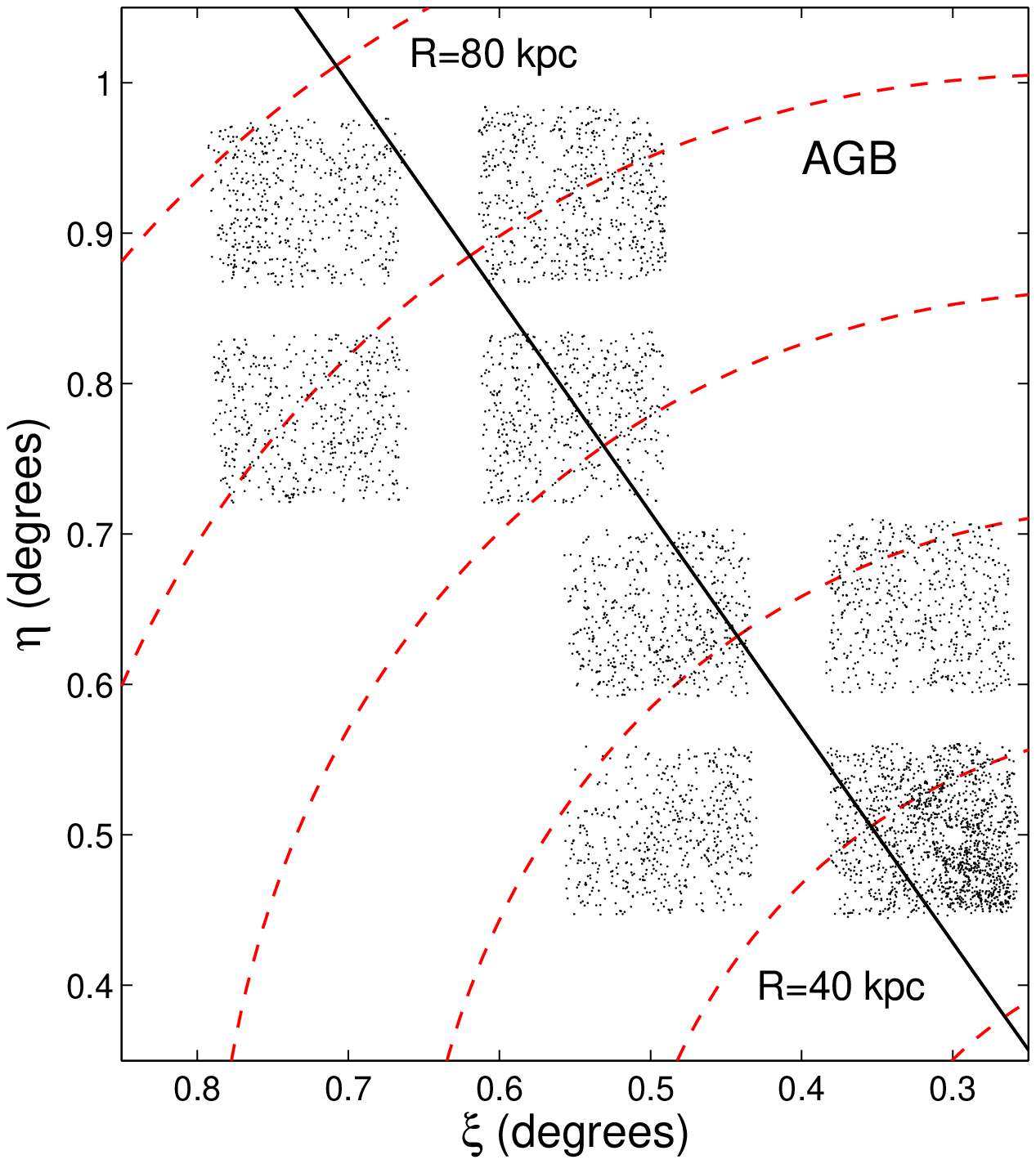}
 \includegraphics[width=8.cm]{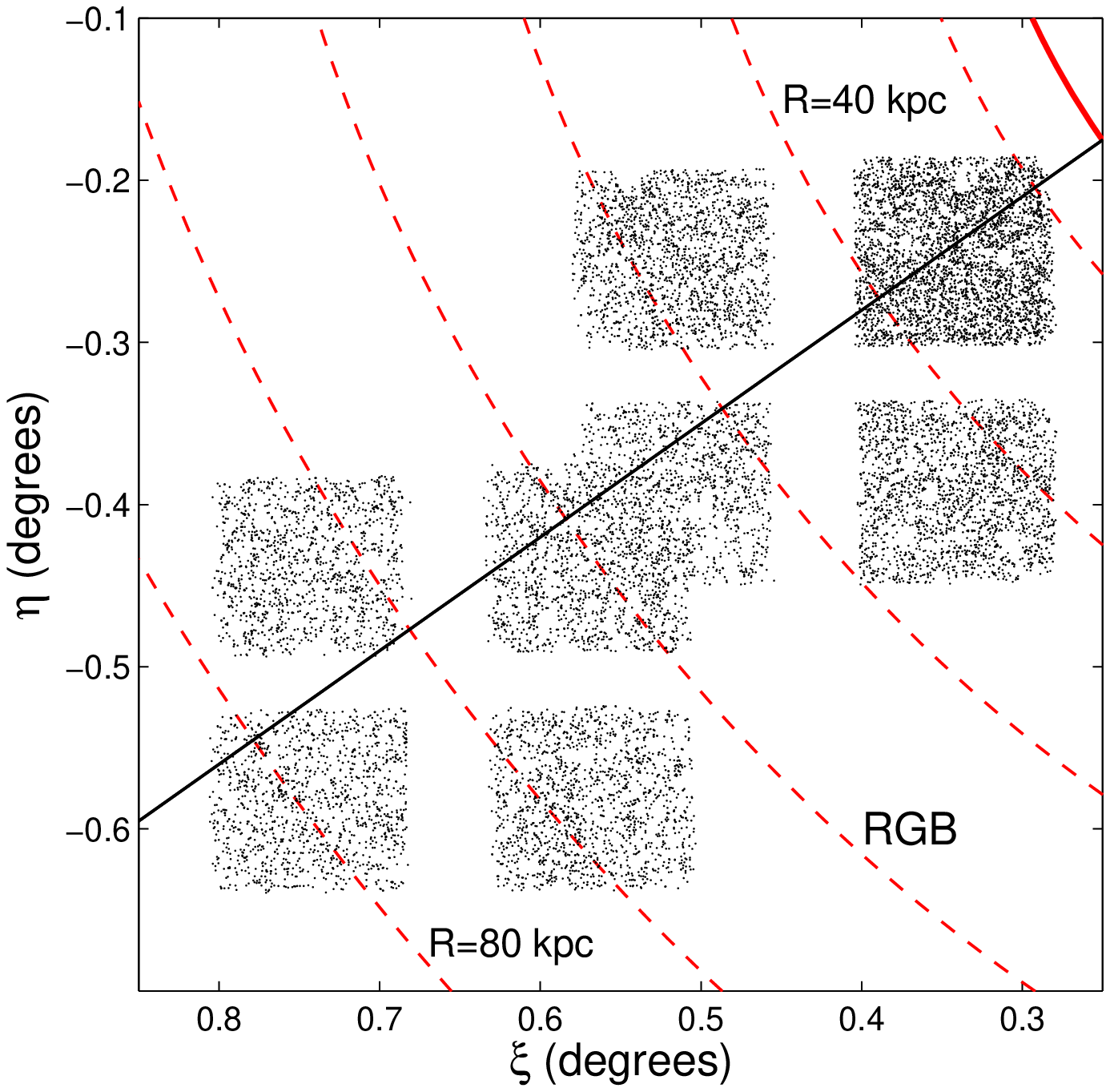}
 \includegraphics[width=8.cm]{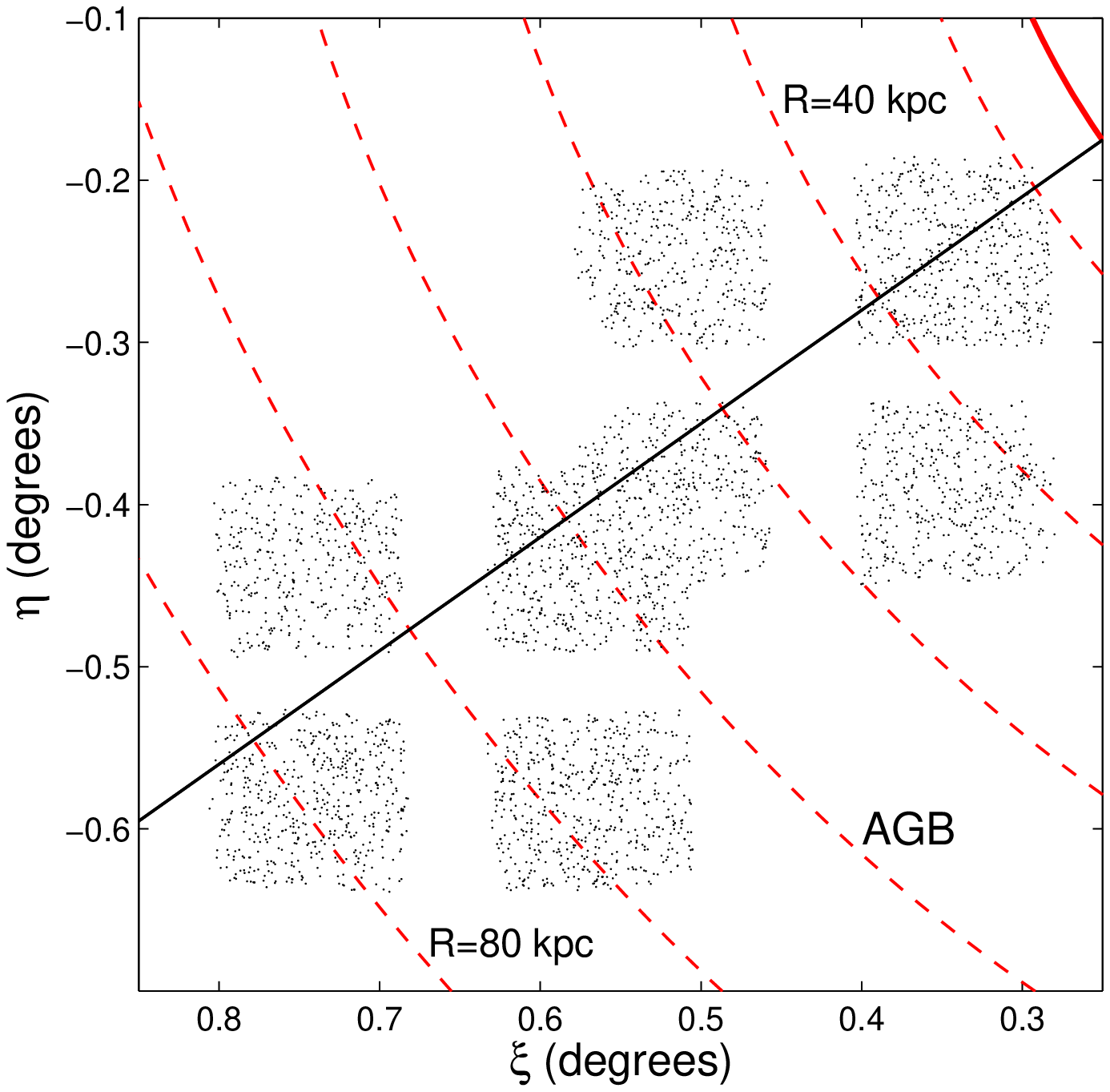}
 \caption{Distribution on the sky (standard coordinates with respect
   to the center of Cen~A) for the candidate RGB and AGB stars along
   the major (\emph{upper panels}) and minor (\emph{lower panels})
   axes (see Fig. \ref{spatdist} for a reference). Red dashed ellipses
   are drawn in steps of 10~kpc, starting at an elliptical
   radius of 30~kpc out to 80~kpc. Black lines are, respectively,
   major and minor axes. Holes in the stellar distributions are mostly
   due to the presence of foreground saturated stars. Note that the
   highest density region is localized to the innermost chip along the
   major axis, for both RGB and AGB stars. To underline the highest
   density in the RGB sample, we replace individual points with a 
   contour plot for densities larger than $\sim670$~stars/arcmin$^2$.}
\label{onsky}
\end{figure*}

\begin{figure}
  \centering
 \includegraphics[width=8.cm]{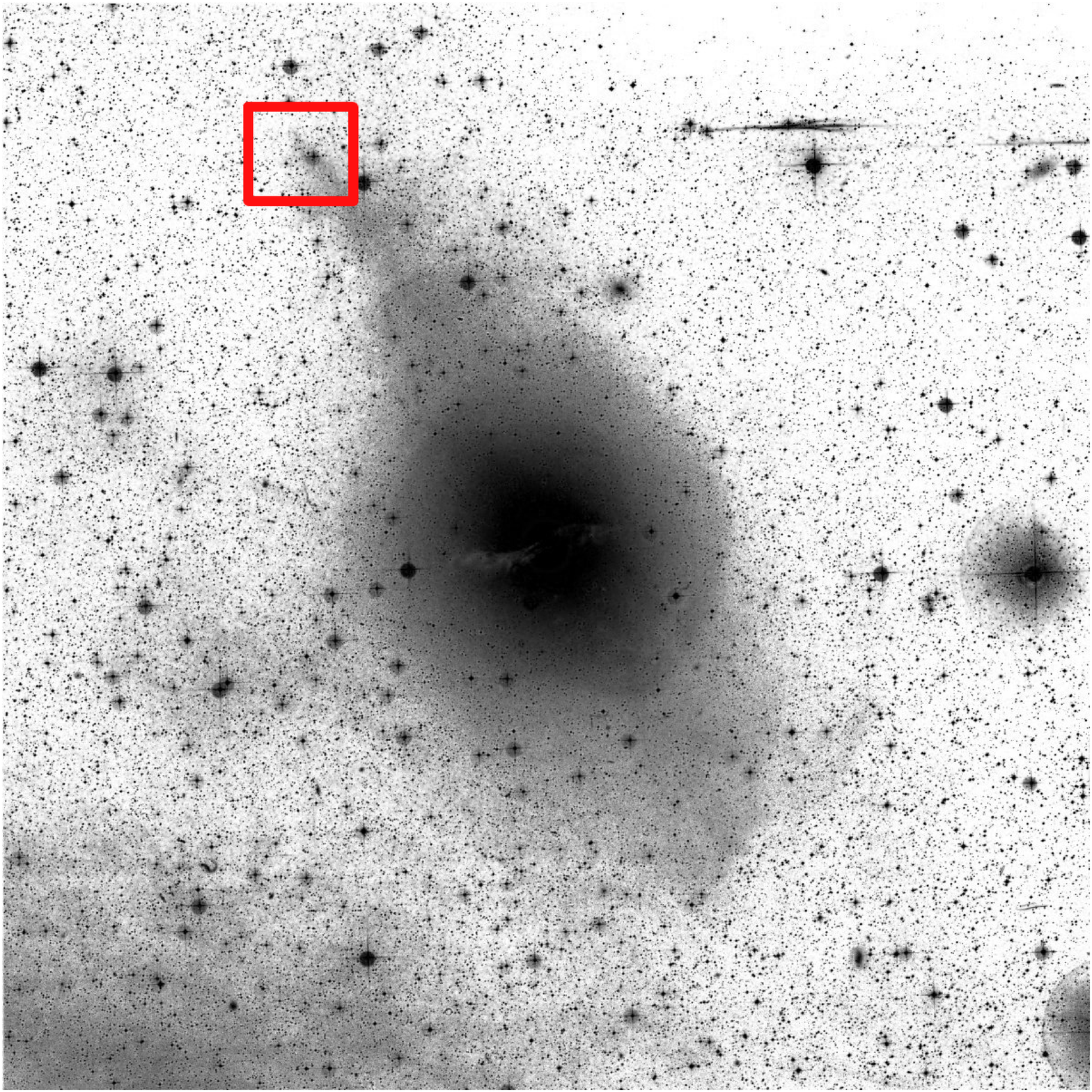}
 \caption{Extended UKST photographic plate image (see also Fig.
   \ref{spatdist}; north is up, east is left).  Overlaid is the
   location of the innermost VIMOS chip along the major axis (red
   rectangle).}
\label{photplate}
\end{figure}

In Fig. \ref{onsky} we show the spatial distribution of sources
falling within our RGB and AGB selection boxes.  Small holes in the
distribution are mostly due to saturated stars. As noted before, the
highest stellar density occurs in the maj1 pointing.  Additionally,
Fig. \ref{onsky} reveals that the enhanced density in this region is
due in part to a spatially-coherent substructure which is most
prominent on the innermost chip; this overdensity is apparent in both
the distribution of RGB and AGB candidates.  At a radius of $\sim
40$~kpc, the new substructure lies well beyond the known shell system.
To further explore this feature, we retrieved a scanned UK Schmidt
Telescope (UKST) IIIa-J survey plate centred on Cen~A and spanning
$1.4^{\circ}$ on a side.  Fig. \ref{photplate} shows a slightly
smoothed version of this plate with an outline of the innermost chip
of the maj1 pointing overlaid.  The substructure detected in our star
count analysis is clearly visible on this deep plate.  Furthermore,
the plate reveals that the substructure is part of a very large
overdensity of stars which protrudes from the main body of Cen~A along
the north-eastern major axis.  Given the morphology of this feature,
the most likely origin is post-merger debris from a significant merger
event and not individual dwarf galaxy accretions.
  
\begin{figure*}
  \centering
\includegraphics[width=8.cm]{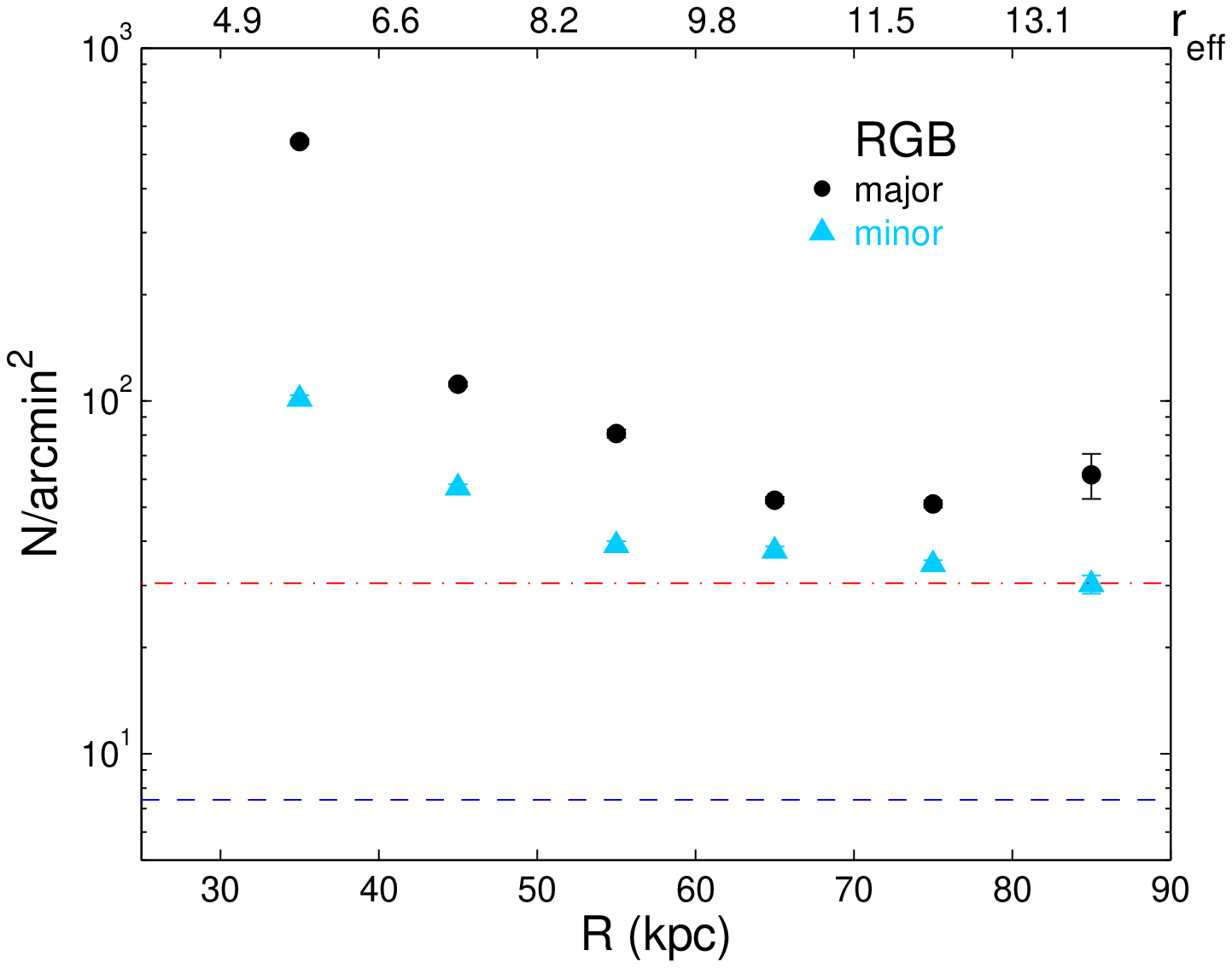}
 \includegraphics[width=8.cm]{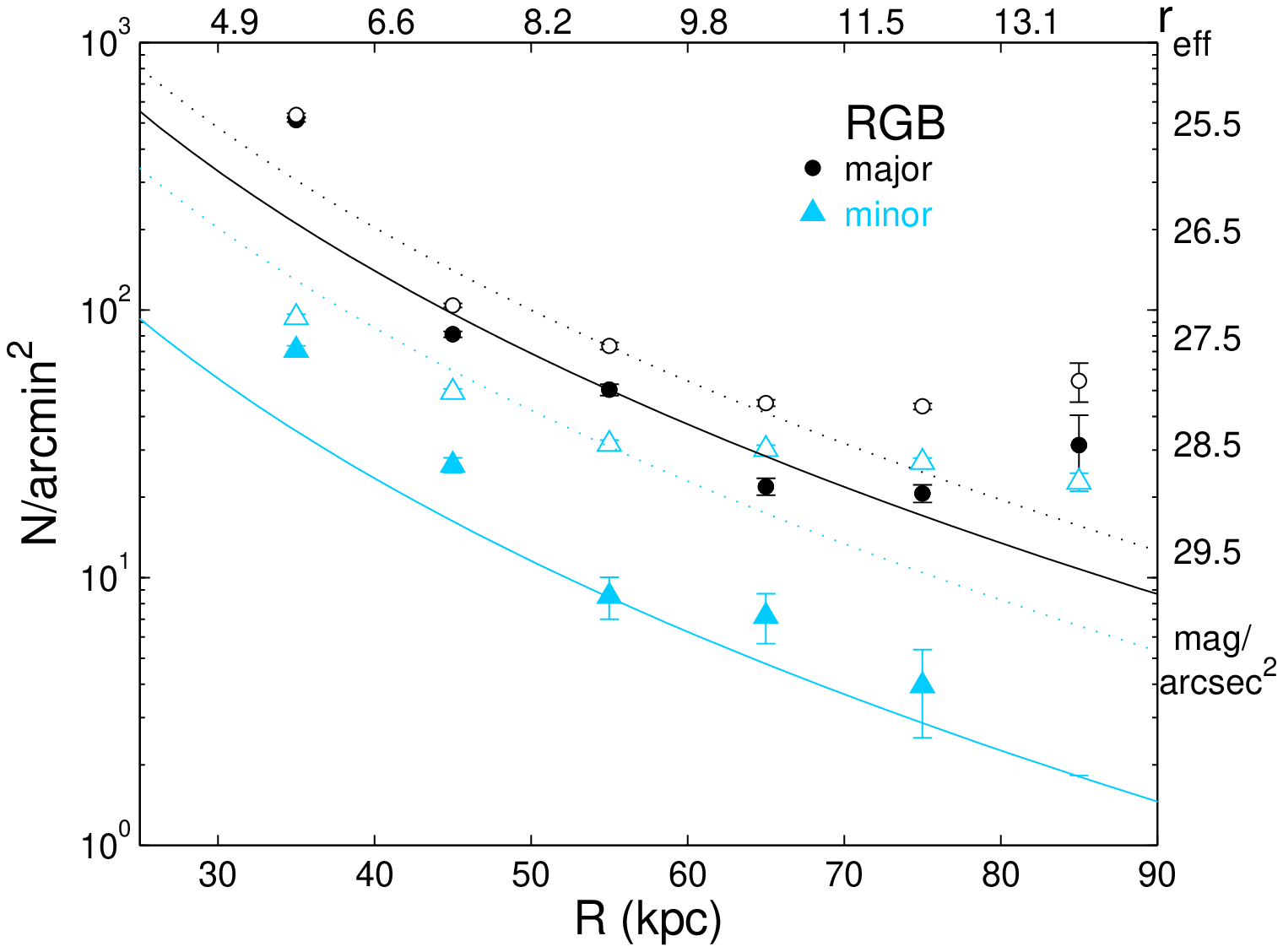}
 \caption{\emph{Left panel}. Radial density profiles for RGB stars
   along both major (filled black circles) and minor (filled light
   blue triangles) axes, as a function of elliptical radius.
   The radius is also expressed in terms of effective radii (top
   axis). Star counts have been corrected for incompleteness (both
   radial and in magnitude). The dashed-dotted red line
   indicates the upper limit estimate for background/foreground
   contamination (i.e., the RGB number density for the $75<R<85$~kpc
   bin along the minor axis), while the dashed blue line represents
   the lower limit estimate (i.e., the number density of resolved
   background sources and simulated foreground stars). 
   \emph{Right panel}. Same as the left panel, after subtraction of
   the upper limit estimate (filled symbols) and lower limit estimate
   (open symbols) for background/foreground. Overplotted for each axis
   and for each of the two subtracted profiles (solid lines for the
   upper limit and dashed lines for the lower limit for
   background/foreground) is a de Vaucouleurs' profile with effective
   radius of 6.1~kpc and arbitrarily scaled to match the datapoint at
   $R=55$~kpc. The right-hand vertical axis reports surface brightness values 
   corresponding to the de Vaucouleurs' profile arbitrarily scaled 
   to the (upper-limit) contaminant-subtracted major axis profile.
   The errorbars are poissonian.}
\label{densprof}
\end{figure*}

Aside from the innermost major axis field, the RGB stellar density
declines smoothly with radius along both axes, while for candidate AGB
stars it remains relatively constant.  We show the radial density
profiles for the RGB sample in the left panel of Fig. \ref{densprof}.
These are computed using the same elliptical annuli as before.  Star
counts are calculated in radial bins for $30<R<90$~kpc, in steps of
10~kpc, and corrected for incompleteness as a function of radius and
magnitude.  These profiles are not corrected for foreground/background
contaminants.  Instead, we indicate the RGB number density for the
$75<R<85$~kpc bin along the minor axis with a dash-dotted line, and
assume that this represents an upper limit on the contamination.
Similarly, a lower limit is given by the dashed line which is
constructed from the sum of resolved background source density and
simulated foreground star density, derived in the same selection box
as for the RGB sample.  As can be seen, these contaminant estimates
differ by a factor of $\sim4$.  Completeness corrections are very
significant for $R<40$~kpc while the last major axis point is
calculated from only a very small area (see Fig \ref{onsky}).  It is
clear that the surface density of RGB stars is always higher along the
major axis than the minor axis, and especially for radii
$\lesssim55$~kpc.  On the other hand, both axes exhibit similar
profile shapes with steep declines in the number counts out to $\sim
55$~kpc and much flatter declines beyond that.

Unfortunately, we cannot construct a surface density (or surface
brightness) profile across the whole extent of Cen~A's halo since the
UKST plate does not have the required level of linearity for an
accurate photometric calibration.  Previous studies have constrained
the structure of Cen~A over small areas. Using data within the inner
$\sim12$~kpc,  \citet{vandenbergh76} fit a de Vaucouleurs' law \citep{devauc59} with parameters $R_{\rm
  eff}=6.1$~kpc and $V(R_{\rm eff})=22.15$~mag arcsec$^{-2}$.
Additionally, \citet{dufour79} find $R_{\rm eff}=5.5$~kpc from $U$-
and $V$-band profiles along the major axis, with $V(R_{\rm
  eff})=22.00$~mag arcsec$^{-2}$. The smaller effective radius found
in the latter study probably results from their exclusion of data at
radii $\gtrsim4.7$~kpc.

\begin{figure*}
  \centering
 \includegraphics[width=8.5cm]{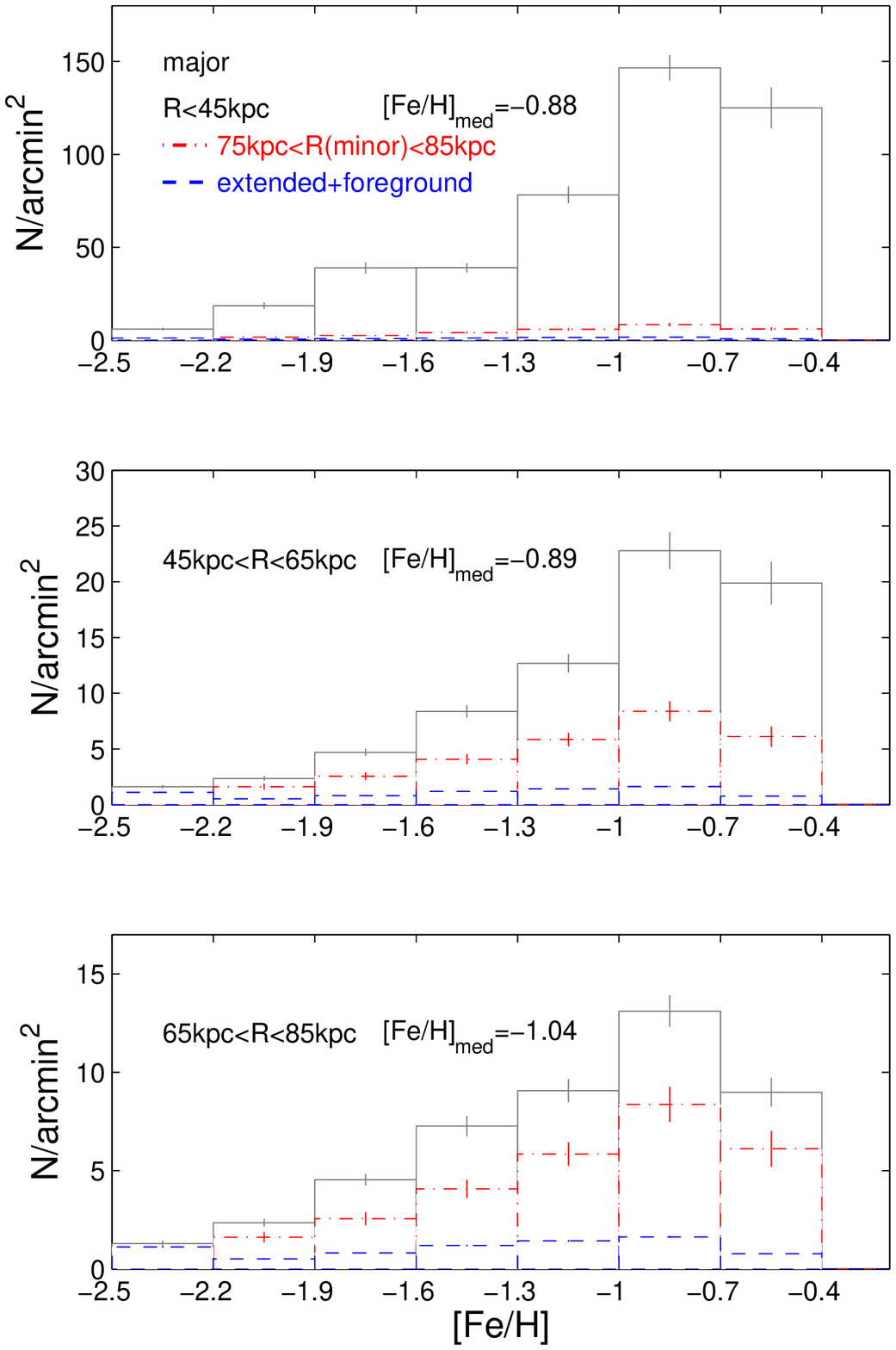}
 \includegraphics[width=8.5cm]{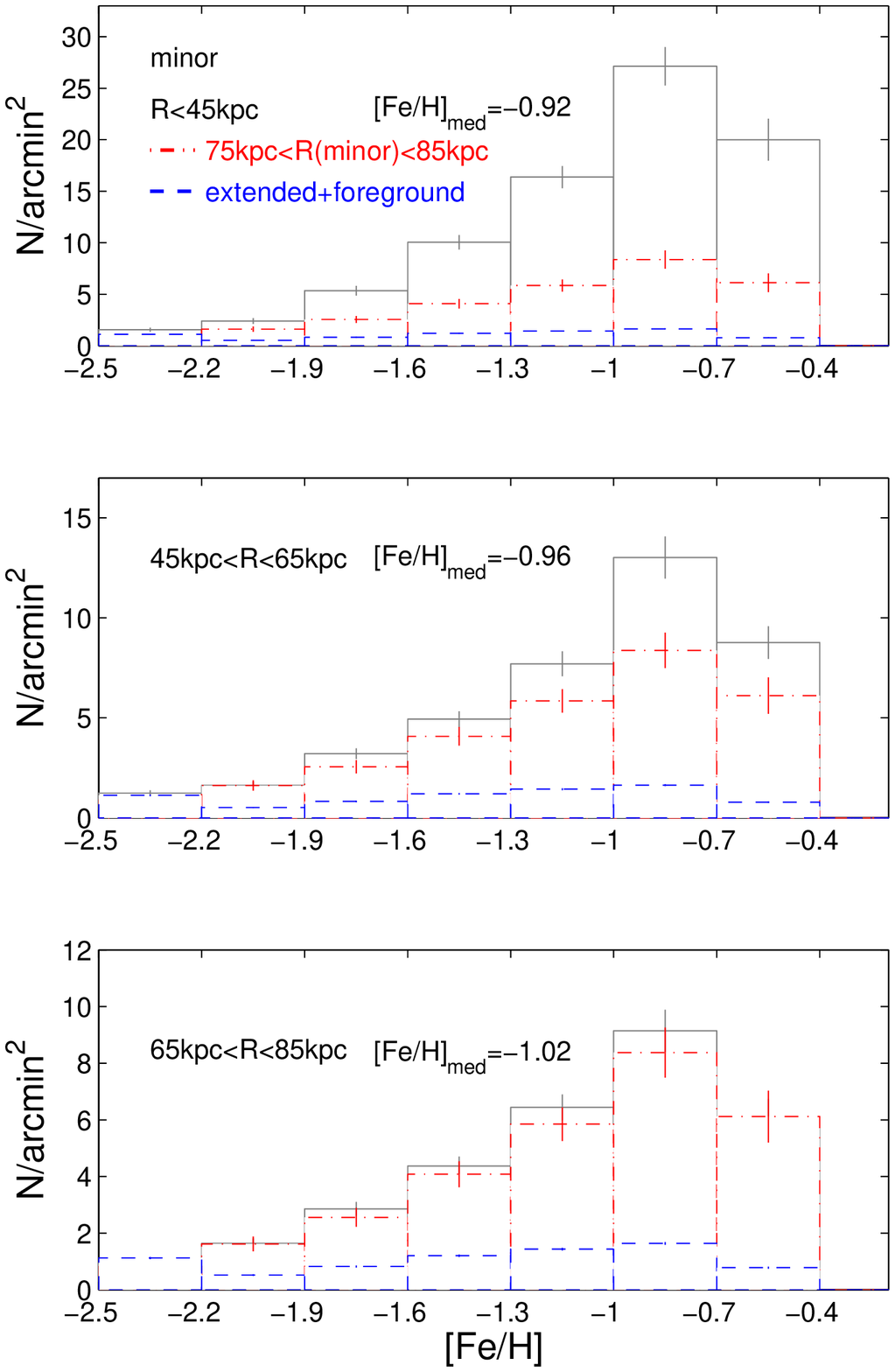}
 \caption{Metallicity distribution functions as a function of
   elliptical radius, for both major (\emph{left panels})
   and minor (\emph{right panels}) axes (solid black lines).  The MDFs
   are derived via isochrone interpolation with a fixed age (12 Gyr)
   and varying metallicity. The histograms are corrected for
   incompleteness and normalized per unit area, and the median
   metallicity (after completeness correction) is reported for each
   panel.  The errorbars are computed via MC simulations, taking into
   account photometric errors and random errors stemming from the
   interpolation technique. Finally, in each subplot we also show the
   MDF for the $75<R<85$~kpc bin along the minor axis (red dot-dashed
   histograms, i.e., upper limit estimate for the
   foreground/background contamination), and the combined MDF for
   resolved background sources and simulated foreground stars (blue
   dashed histograms, i.e., lower limit estimate).}
\label{spatmdfs}
\end{figure*}

It is of obvious interest to explore whether the stellar density
profile in the outer halo behaves as expected from an extrapolation of
the inner profile and we attempt to address this as follows.  In the
right panel of Fig. \ref{densprof}, we present contaminant-subtracted
versions of our RGB radial density profiles using both the upper
(filled symbols) and lower (open symbols) limit estimates for the
contaminant level.  We overplot the de-reddened de Vaucouleurs'
profile derived by \citet{vandenbergh76}, arbitrarily scaled to match
the $R=55$~kpc datapoint in each case.  Our VIMOS density profiles
appear broadly consistent with fall-off of the extrapolated de
Vaucouleurs' profile.  There is a reasonable agreement along the minor
axis when the upper limit on the contaminant level is adopted, while
using the lower limit yields an excess of stars at radii
$R\gtrsim65$~kpc.  The major axis profile deviates significantly from
the inner extrapolation at $R\geq75$~kpc regardless of what
contaminant level is subtracted. Although some part of this behaviour 
could be due to a genuine flattening of the halo profile at these radii, 
the increased radial density in the $R\sim85$~kpc bin along the major axis 
suggests additional low-level substructure in these parts. 
Visual inspection of the outermost chip along the major axis (Fig. \ref{onsky}) provides a 
tantalising suggestion of this, with a marginally higher stellar density in the outer  
portion of the chip. Had we assumed a de Vaucouleurs'
profile with an effective radius of 5.5~kpc, as derived by
\citet{dufour79}, our conclusions would be unchanged.  For reference,
in the right panel of Fig. \ref{densprof} we indicate the
surface brightness values corresponding to the extrapolated de Vaucouleurs' profile,
arbitrarily scaled to match the (upper-limit) contaminant-subtracted major
axis profile at 55~kpc.

Finally, the radial density profiles for candidate AGB stars (not
shown here) are essentially flat along both axes, except for the
significant major axis overdensity at radii $\lesssim55$~kpc. Any
genuine AGB stars in these parts are most likely completely
outnumbered by foreground and background contaminants.


\section{Metallicity Distribution Functions} \label{mdfs_sec}

For largely coeval populations, the colour and width of the RGB
constrains the mean metallicity and metallicity spread of the
population (the RGB colour is more sensitive to changes in metallicity
than it is to changes in age, \citealt[e.g.,][]{vandenberg06}.). Given
that we have not found evidence for a significant intermediate-age AGB
population in our VIMOS fields (except for the innermost radial bin
along the major axis), we will assume that the Cen~A halo population
is essentially coeval and proceed to derive photometric MDFs
\citep[see, e.g.,][and references therein]{harris99, harris02,
  rejkuba05}.

We again adopt 12~Gyr $\alpha$-enhanced ([$\alpha$/Fe] $=+0.2$)
isochrones from the Dartmouth group \citep{dotter08}, with metallicity
ranging from [Fe/H] $=-2.5$ to [Fe/H] $=-0.3$, sampled every 0.2 dex.
This choice of age is driven by the analysis of \citet{rejkuba11}
which indicates that the bulk of the star formation in their 38~kpc
field occured $\sim12$~Gyr ago on a relatively short timescale
($\sim3$~Gyr).

Our MDFs are calculated only using stars which fall within our RGB
selection box, to ensure good completeness.  In this region of the
CMD, the stellar isochrones are also more widely separated thus
reducing the uncertainty in individual metallicity determinations. We
linearly interpolate within the isochrone grid to determine the
metallicity of each star. The uncertainties are computed by producing
1000 Monte-Carlo (MC) realizations of the interpolation process: each
time the CMD position of the star is varied within its photometric
uncertainty (with a Gaussian distribution), and the metallicity
recomputed.

While the derivation of precise \emph{absolute} metallicity estimates
is not possible with photometric data alone, we can nevertheless get
robust results on the \emph{relative} metallicities within our sample.
Fig. \ref{spatmdfs} shows the MDFs, corrected for incompleteness, as a
function of radius along both axes, while Tab. \ref{metdiff} tabulates
their main properties.  The MDFs exhibit the same overall shape
consisting of a peak around [Fe/H]$\sim-0.8$ and a long tail to lower
metallicities.  At $R<65$~kpc, the major axis has a slightly larger
number of stars more metal-rich than [Fe/H]$\sim-0.70$ compared to the
minor axis, which translates into a small increase in the median
metallicity.  This mild enhancement could plausibly be related to the
presence of the substructure we have uncovered in these parts.
Intriguingly, we find the median [Fe/H] is rather constant across the
radial extent of our survey, varying from only $-0.88$ to $-1.04$.
In particular, there is a change of $\Delta$[Fe/H]$=-0.16$~dex along
the major axis ($\sim35-80$~kpc, see Fig \ref{onsky}) corresponding to
a gradient of $\sim-0.004$~dex/kpc.  Along the minor axis, the change
is $\Delta$[Fe/H]$=-0.1$~dex over $\sim30-85$~kpc and thus a similarly
small gradient of $\sim-0.002$~dex/kpc.  The metallicity spread, as
estimated by the standard deviation of the individual metallicity
measurements, is also essentially constant at $\sim0.45$~dex
throughout the halo.  Finally, the fraction of metal-poor stars (which
we define as $f_{mp}=$ N$_{[Fe/H]<-1.0}$/N$_{tot}$) shows only a very
modest increase of $\sim10\%$ with radius, reaching a maximum of
$\sim50\%$ in the outermost bin.

 \begin{table*}
 \begin{minipage}{170mm}
  \centering
   \caption{Median metallicities and metal-poor stars fractions ($f_{mp}=$ N$_{[Fe/H]<-1.0}$/N$_{tot}$) as a function
     of radius, for different background/foreground subtractions.}
 \label{metdiff}
   \begin{tabular}{lcccccc}
     \hline
     \hline
     Field & [Fe/H]$_{med}$ & [Fe/H]$_{med,sub-upp}$\footnote{after
       subtraction of the upper limit on background/foreground estimate,
       i.e., the MDF of the $75<R<85$~kpc bin along the minor axis.} &
     [Fe/H]$_{med,sub-low}$\footnote{after subtraction of the lower
       limit estimate, i.e., the combined MDF of resolved background
       sources and simulated foreground stars.} &
     $f_{mp}$ & $f_{mp,sub-upp}$ & $f_{mp,sub-low}$ \\
     \hline
     {$R<45$~kpc (major)}&-0.88$\pm0.04$&-0.87$\pm0.04$&-0.87$\pm0.04$&0.40$\pm0.04$&0.39$\pm0.05$&0.39$\pm0.05$\\
     {45~kpc $<R<65$~kpc (major)}&-0.89$\pm0.01$&-0.85$\pm0.01$&-0.87$\pm0.01$&0.41$\pm0.01$&0.34$\pm0.01$&0.38$\pm0.01$\\
     {65~kpc $<R<85$~kpc (major)}&-1.04$\pm0.01$&-1.08$\pm0.05$&-1.00$\pm0.01$&0.53$\pm0.01$&0.56$\pm0.04$&0.50$\pm0.01$\\
     \hline
     {$R<45$~kpc (minor)}&-0.92$\pm0.01$&-0.89$\pm0.01$&-0.90$\pm0.01$&0.43$\pm0.01$&0.39$\pm0.01$&0.41$\pm0.01$\\
     {45~kpc $<R<65$~kpc (minor)}&-0.96$\pm0.01$&-0.87$\pm0.03$&-0.91$\pm0.01$&0.46$\pm0.01$&0.34$\pm0.04$&0.41$\pm0.01$\\
     {65~kpc $<R<85$~kpc (minor)}&-1.02$\pm0.01$&-1.09$\pm0.27$&-0.96$\pm0.01$&0.52$\pm0.01$&0.62$\pm0.22$&0.47$\pm0.01$\\
     \hline
     \hline
 \end{tabular}
 \end{minipage}
 \end{table*}
 
\begin{figure}
  \centering
 \includegraphics[width=8.cm]{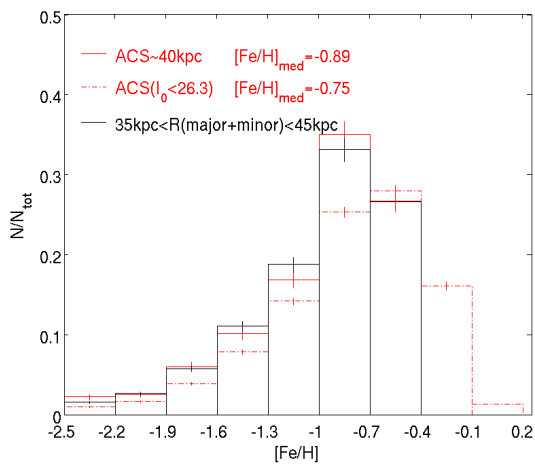}
 \caption{Normalized metallicity distribution functions for the
   combined VIMOS fields at $35<R<45$~kpc along major and minor axes
   (black solid line), compared to the one for the outermost ACS/HST
   pointing (red solid line).  The solid line MDFs are drawn from
   stars in the RGB selection box shown in Fig. \ref{cmdsfore}, and
   the VIMOS MDFs have been corrected for incompleteness. The red
   dot-dashed MDF is derived using all RGB stars brighter than
   $I_0=26.3$ and within the whole colour range of the ACS/HST
   photometry. The VIMOS errorbars are drawn from MC simulations (see
   text), while the ACS errorbars are poissonian.}
\label{mdfscomp}
\end{figure}

\subsection{Uncertainties}

There are a number of uncertainties in the derivation of the MDFs that
we now consider.  In particular, we wish to address to what extent
these uncertainties could affect our derived MDFs as a function of
radius. Firstly, intermediate-age AGB stars ascending the giant branch
could contaminate our RGB samples and this could be important in our
innermost major axis chip where the clear signature of such a
population exists.  AGB stars will lie on the blue side of the RGB,
and will thus artificially enhance the metal-poor population in our
MDFs.  However, given that the metal-poor population in the innermost
major axis bin is small (see Fig. \ref{spatmdfs}), we conclude that
this cannot be a significant effect.

Secondly, similar to the LFs discussed in Sec. \ref{lfs_sec}, the MDFs
we derive will contain both genuine Cen~A halo stars as well as
contaminants.  As the contaminants are not RGB stars at the distance
of Cen~A, the metallicities computed for them will be meaningless.
Both the number and the colour distribution of the contaminants can
potentially influence our results.  In Fig. \ref{spatmdfs}, we show
normalized ``MDFs" for our two estimates of the contaminant population,
derived in exactly the same way as described above.  Our lower limit
estimate, based on summing resolved galaxies and foreground models,
yields an ``MDF" that is fairly flat while our upper limit estimate,
based on the outermost portion of the outer minor axis field, bears a
strong similarity to those we have derived for the halo. This is not
surprising since this field may still contain a significant number of
genuine Cen~A RGB stars.  We subtract these estimates of the
contaminant ``MDF" from our measured MDF and recompute the median
metallicity and metal-poor fractions (see Tab. \ref{metdiff}).  This
has only a small impact on our results and does not affect our
inferences regarding the metallicity gradient and metal-poor fraction.

Finally, although we have corrected for incompleteness in deriving our
MDFs, these corrections are very significant for the reddest
($(V-I)_0\gtrsim2.5$) populations that fall within our RGB selection
box. In order test how accurate our completeness corrections are for these
metal-rich populations, we use the photometric catalogue for the
HST/ACS field at 38~kpc, kindly provided by M. Rejkuba
\citep{rejkuba05}, which reaches to redder colours and fainter
magnitudes than our VIMOS data. We deredden the photometry with the mean
extinction values reported in Sect. \ref{cmds}.  For the magnitude and
colour ranges of relevance for comparing to our VIMOS data, the
photometric and colour uncertainties are generally smaller than
$\sim0.03$ and $\sim0.04$ mag respectively, and the completeness as
derived from artificial star tests is close to $\sim100\%$.  In
addition to photometric accuracy, the other main advantage of this
catalogue over our ground-based data is its purity.  The high
resolution (0.049 arcsec/pixel) of the ACS means that background
galaxy contaminants can easily be identified and rejected.

We initially derive the MDF for the HST/ACS catalogue using the same
RGB selection criteria and procedure as for our VIMOS data.  In Fig.
\ref{mdfscomp} (solid lines), we show the resulting ACS MDF as well as
the VIMOS MDF for the same radial range ($35<R<45$~kpc), constructed
using data along both major and minor axes.  The HST/ACS sample yields
a median metallicity of [Fe/H] $=-0.89$, and a spread (standard
deviation) of $\sim0.44$~dex. For a comparison, the VIMOS MDFs at
these radii have a median metallicity of [Fe/H] $=-0.90$ with a spread
of $\sim0.42$~dex. Reassuringly, there are only small differences in
the median metallicity and the shape of the two MDFs, confirming that
that our completeness corrections for the reddest stars in the VIMOS
data are accurate, and that contaminating galaxies have a negligible
influence on our results at these radii.

We then proceed to derive the MDF for the HST/ACS data using a broader
range of colour and magnitude than we are sensitive to with VIMOS in
order to assess what systematic biases might arise from the
limitations of our photometry.  We select RGB stars brighter than
$I_0=26.3$ and use the whole range of colours in the HST/ACS CMD.  The
dashed line in Fig. \ref{mdfscomp} shows the resultant MDF in this
case.  Overall, there is a good agreement between this MDF and our
VIMOS MDF however it peaks at somewhat higher metallicities and has a
metal-rich tail not seen in the ground-based data. The median
metallicity derived in this way is [Fe/H] $=-0.75$ with a spread of
0.46~dex.  This value translates into [M/H] $=-0.61$ for our choice of
$\alpha$-enhancement, and is in excellent agreement with the mean
metallicity [M/H] $=-0.65$ reported by \citet{rejkuba05} (obtained
using the \citealt{vandenberg00} evolutionary models). Finally, the
metal-poor fractions for the HST/ACS MDFs are 0.38 (VIMOS selection
box) and 0.29 (full colour range) which are within $\sim10\%$ of our
VIMOS estimate (i.e., 0.40 for the considered radial range).

In summary, none of the effects discussed above are expected to have a
significant impact on our results. The lack of sensitivity to very red
stars in the VIMOS data may slightly bias ($\sim 0.15$~dex) our
metallicity estimates towards lower values, but it is unlikely that
this effect could disguise any strong radial gradients. Contamination
from background/foreground objects has even less of an effect on our
measurements.  We conclude that the moderately high and relatively
uniform metallicity in the extended halo of Cen~A is a robust result.


\section{Discussion} \label{disc}

\subsection{Halo Extent and Structure}

The intrinsic faintness of the outskirts of gEs has made it
challenging to quantify the structure and extent of these objects,
even in our local universe. We have shown the existence of an RGB
population to at least $\sim85$~kpc ($\sim14 R_{\rm eff}$) along the
major axis of Cen~A.  Along the minor axis, there is a clear RGB
signature in the CMDs to $\sim65$~kpc, however analysis of LFs and
TRGB distances suggests that the stellar population extends further in
this direction too.  The RGB population in Cen~A is therefore at least
as extended as the distribution of planetary nebulae and GCs
\citep{peng04, peng04b, woodley07, woodley10, harris12}.  A few
studies have shown the existence of stars at very large radii in other
ellipticals (e.g., \citealt{weil97,harris07a,harris07b,tal11}).  Taken
together, these studies demonstrate the vast extents of elliptical
galaxies.  Almost all previous studies of ellipticals have focused on
regions lying at $\lesssim 1-2 R_{\rm eff}$. While this is sufficient
to sample the bulk of the light in these systems, it samples only a
small fraction of their physical extents.  With the increasing focus
on questions related to the size evolution of early-type galaxies
(e.g., \citealt{buitrago08}), an understanding of the properties and
nature of the peripheral regions of ellipticals is of paramount
importance.

The RGB stellar density in Cen~A's halo is always higher along the
major axis than the minor axis for a given elliptical annulus.  Within
$\sim55$~kpc ($<9 R_{\rm eff}$), this results from the presence of a
coherent overdensity of stars that we have uncovered along the major
axis.  The morphology of this feature suggests it is post-merger
debris, perhaps originating from the same recent ($\lesssim1$~Gyr)
event that led to the galaxy's peculiar inner structure \citep[see,
e.g.,][]{israel98}.  The feature is unlikely to be due to triggered
star formation from Cen~A's radio jets given the estimated timescale
($10^{7-8}$~years) for this process compared to the typical
$\gtrsim1$~Gyr age of RGB stars (\citealt{israel98}).  The existence
of such substructure in Cen~A's far outer halo indicates that these
regions are significantly inhomogeneous and that a wide-field approach
is required to properly study and interpret the resolved stellar
populations in these parts.

Beyond 55~kpc, the difference between the major and minor axis density
profiles could plausibly be explained if our assumption of elliptical
isophotes with b/a=0.77 and a position angle of $35^{\circ}$ breaks
down.  Indeed, the two profiles could be brought into better agreement
if b/a$\sim$0.5-0.6 in these parts, implying a higher ellipticity in
the outer halo.  Interestingly, \citet{tal11} find the same result in
their stacking analysis of SDSS luminous red galaxies. In particular,
they find a mild increase in ellipticity from 0.25 to 0.3 (or
equivalently b/a varying from 0.75 to 0.7) at radii $\gtrsim5 R_{\rm
  eff}$.  On the other hand, the profile differences in Cen~A could
simply be due to additional low-level substructure further out along
the major axis, a tantalising hint of which may exist in our data.

Finally, we have examined whether the star count measurements we have
made in the outer halo are consistent with an extrapolation of the
inner surface brightness profile.  Along both the major and minor
axes, there is evidence for a flattening in the profile shape beyond
$\sim$60-70~kpc ($\gtrsim~10-11.5 R_{\rm eff}$), the exact details of
which depend on the contaminant level subtracted and how the profiles
are normalised.  Subtracting the upper limit on the contaminant level,
approximate power-law profiles of $R^{-3.3}$ and $R^{-3.8}$ can be fit
to the major and minor axis profiles of Cen~A, respectively. A
flattening of the outer profile was also seen beyond $\sim8 R_{\rm
  eff}$ in the stacked surface brightness profile of \citet{tal11}.
These authors were not able to tell whether the excess light at large
radius was due to a flattening in the galaxy surface brightness
profile, or to intragroup light. Indeed, the same ambiguity persists
to some extent here although the Cen~A group is unlikely to have an
extensive intragroup light component.

\subsection{Metallicity Gradient}

The photometric MDFs derived from our VIMOS data yield median values
of [Fe/H]$_{med} \sim-0.9$ to $-1.0$~dex which, given our choice of
$\alpha$-enhancement, correspond to [M/H]$_{med} \sim-0.75$ to
$-0.85$~dex.  The median metallicity is essentially constant, varying
by only $\sim 0.1-0.15$~dex across the radial extent of our survey
($\sim5$ through $14 R_{\rm eff}$).  This variation implies 
negligible gradients, $\lesssim -0.004$~dex/kpc.  There are also no
strong azimuthal variations, aside from a small enhancement in
metallicity in the region of the outer halo major axis substructure.
This suggests that either the stellar populations of the accreted
system do not differ substantially from the native populations in the
outer halo of Cen~A or that significant mixing of the debris has
already occurred.

The moderately high metallicity found argues against a scenario in
which most of the outer halo stars have been accreted from very low
mass objects. \cite{crnojevic10} study a sample of dwarf satellites of
Cen~A with luminosities in the range M$_{V}\sim -10.7$ to $-13.9$,
corresponding to $\sim 0.005-0.1$\% the luminosity of Cen~A.  They
find median metallicities of [M/H] $\lesssim-1.0$, considerably below
the outer halo metallicity of Cen~A.  However, this result may not be
surprising. Indeed, \citet{oser12} use simulations to argue that the
dominant mode of halo assembly is through minor mergers with typical
mass ratios of $\approx$~1:5; such systems are likely to be more
metal-enriched than low mass dwarfs.

We can compare our results to previous HST studies of Cen~A's halo.
These cover 4 fields in the radial range from $\sim 1.3$ to $\sim
6.2 R_{\rm eff}$, none of which lie along the major axis
\citep{harris99,harris00,harris02,rejkuba05}.  Aside from the central
pointing which contains a combination of halo and bulge stars, these
fields all exhibit a moderately high mean metallicity ([M/H]$\sim
-0.4$ to $-0.6$) with only a small population ($\sim 10$\%) of low
metallicity stars.  When combined with our results presented here,
this suggests a variation of only $\lesssim 0.5$~dex in [M/H] over the
entire extent of the halo.  Harris {\it et al.} were able to reproduce
their MDFs with a simple chemical evolution model in which early star
formation proceeds with an initial stage of rapid infall of very
metal-poor gas, after which the infall dies away exponentially.
Similarily, \citet{rejkuba11} argue that most of the stars in their
field at $\sim 6.2 R_{\rm eff}$ formed rapidly and at very early
times, with a minority intermediate-age component having formed more
recently. The moderate metallicities we find at larger radii in
the halo suggests that the qualitative conclusion of rapid collapse
continues to hold in these parts, although we have yet to model our
MDFs in detail. Interestingly, recent kinematic and abundance studies
of GCs in Cen~A's halo (out to $\sim40$~kpc) also support a fast,
early formation phase, coupled with subsequent merger events
\citep{woodley09, woodley10}.

Unfortunately, it is not straightforward to compare our results to
those for other gEs.  Very few observational studies have been able to
investigate the metallicity content of stellar populations at
galactocentric radii as large as we have probed in Cen~A.  HST/ACS
observations have been obtained centred at $\sim12 R_{\rm eff}$ for the
nearby ``classical'' gE NGC3379 ($\sim11$~Mpc; \citealt{harris07a})
and at $\sim3 R_{\rm eff}$ for the intermediate-sized elliptical, NGC3377 ($\sim10$~Mpc;
\citealt{harris07b}). The resulting picture is different for these two
systems. NGC3379 shows an extremely broad and flat MDF, with the
metal-poor population becoming increasingly dominant at $\gtrsim10 R_{\rm
  eff}$.  On the other hand, NGC3377 has an MDF peaked at $\sim
-0.6$~dex and does not show a significant gradient out to the furthest
radius probed ($\sim5 R_{\rm eff}$).  In integrated light analyses,
some systems show continuing metallicity gradients out to $\sim
2-4 R_{\rm eff}$ \citep{greene12, weijmans09} while others show
steepening of gradients when traced as far out as $8 R_{\rm eff}$
\citep{labarbera12}.  The stacking analysis of \citet{tal11} yields a
clear colour gradient within the inner $\sim 3 R_{\rm eff}$, with the
populations getting bluer and thus more metal-poor (and possibly
older) with radius, while the color index flattens out at larger
galactocentric distances (out to $\sim 14 R_{\rm eff}$).  Larger
samples will be required to establish what is the typical behaviour
for the metallicity gradient at large radii in gEs.

Comparison to the predictions of theoretical models is similarly
difficult since existing models do not explicitly address the large
radii probed by our study. Theoretical models of gE formation are able
to predict both the presence and the absence of metallicity gradients
at smaller radii.  Monolithic collapse models, in which gas sinks to
the centre of the potential well and is enriched by the first
generations of evolving stars, predict steep radial metallicity
gradients which become more pronounced with increasing galaxy mass due
to gas loss efficiency \citep[e.g.,][]{matteucci84, chiosi02,
  kawata03, kobayashi04}. On the contrary, in hierarchical models, the
physical properties of the resultant galaxy will largely depend on
those of the progenitors however the expectation is for metallicity
gradients to be shallower due to the mixing of stars
\citep{kobayashi04, naab09}.  Whether or not these inner trends should
extend to larger radii remains an open question.
 
 Finally, we note in passing that the metallicity of RGB stars in Cen~A's outer halo
 is very similar to that which we  have found in the extended stellar envelopes
 around some spiral galaxies \citep{barker09,barker11}. These extended envelopes
 are also characterised by flatter surface brightness profiles than seen in the
 inner regions.  It is interesting to speculate whether moderate metallicity extended stellar structures 
 could be an ubiquitous feature of all large galaxies, regardless of morphological type.
  

\section{Conclusions} \label{concl}

We have conducted a wide-field survey of resolved stellar populations
in the outer halo of our closest gE, Cen~A, using the VIMOS imager
mounted on the VLT. Two fields were imaged along each of the major and
minor axes sampling projected elliptical radii in the range
$\sim30-85$~kpc (or $\sim5-14 R_{\rm eff}$). From PSF-fitting
photometry, we derive CMDs in $V$- and $I$-bands that reach
$\sim2$~mag below the TRGB.  The CMDs show evidence for an old RGB
population, which can be traced to the furthest extent of our survey
($\sim85$~kpc) along the major axis and to at least $\sim65$~kpc along
the minor axis.  However, consideration of LFs and MDFs strongly
suggests that the stellar population along the minor axis extends to
at least $\sim 85$~kpc too.

The spatial distribution of RGB stars has been mapped as a function of
radius and azimuth. We have uncovered a prominent localized
overdensity extending to $\sim55$~kpc ($\sim9 R_{\rm eff}$) along the
major axis.  Inspection of a deep UKST plate confirms the existence of
this structure and shows a morphology consistent with post-merger
debris. Our VIMOS data indicate that the metallicity of the
constituent stars in this structure do not differ significantly from
that of the underlying halo population.  Even beyond the extent of
this substructure, the major axis stellar density is higher than the
minor axis one at a given elliptical radius.  This could be evidence
for an increasing ellipticity in the outer halo, or conversely for
residual low level debris contaminating the outer halo major axis.
Uncertainties in the contaminant level make it difficult to rigorously
assess whether the outer halo density profile follows an extrapolation
of de Vaucouleurs law which characterises the inner regions.  Both the
major and minor axis profiles appear generally consistent with such an
extrapolation, but there is evidence for a flattening in the profile
beyond $\sim 60-70$~kpc ($\sim10-11.5 R_{\rm eff}$).

We derive photometric MDFs via isochrone interpolation of RGB stars.
The median metallicities we find are relatively high
($<$[Fe/H]$>_{med} \sim-0.9$ to $-1.0$~dex, corresponding to [M/H]
$\sim-0.75$ to $-0.85$~dex), with broad spreads of $\sim0.45$~dex.  We
analyse the median metallicities as a function of radius in order to
search for radial gradients, but find only a modest decrease of
$\sim0.1-0.15$~dex over a range of $\sim5$ to $14 R_{\rm eff}$.
Moreover, the fraction of stars more metal-poor than [Fe/H] $=-1.0$
increases by only $\sim10\%$ as a function of radius, reaching
$\sim50\%$ at the outermost radii probed.  When combined with
previously published results, the stellar metallicity in Cen~A can be
seen to vary by $\lesssim 0.5$~dex over the entire radial range of
85~kpc.  This apparent constancy of the metallicity is vey intriguing,
and suggests that if Cen~A does have a metal-poor halo component then
it can only dominate much further out.

Our results demonstrate that important constraints on elliptical
galaxy formation can come from studying the extent and properties of
their extended halos.  The faintness and inhomogeneity of these parts
requires an approach that combines both depth and wide-field coverage.
Unfortunately, Cen~A is the only gE that can, at present, be readily
resolved into stars over large areas from the ground; other
techniques, perhaps less optimal, will be required for additional gEs.
Our findings indicate that Cen~A itself is also worthy of further
study; observations at even larger radii than we have probed here will
be required in order to establish its true extent and search for
evidence of an underlying metal-poor component to its stellar halo.


\section*{Acknowledgments}

We thank the service mode support staff at Paranal for conducting the
VIMOS observations.  We kindly acknowledge M. Rejkuba for sharing the
photometric catalogue of her HST/ACS pointing, and N. Hambly for
providing the scanned photographic plate image of Cen~A.  
We thank Rodrigo Ibata for providing useful comments that helped us improve 
our manuscript. DC acknowledges hospitality from the Mullard Space Science Laboratory -
University College of London, where part of this work was carried out.
DC, AMNF and EJB are supported by an STFC Rolling Grant.  This
research made use of the NASA/IPAC Extragalactic Database (NED), which
is operated by the Jet Propulsion Laboratory, California Institute of
Technology, under contract with the National Aeronautics and Space
Administration.

\bibliographystyle{mn2e}
\bibliography{biblio.bib}


\label{lastpage}


\end{document}